	\documentclass{aa}
	\usepackage{txfonts}
	\usepackage{graphicx}
        \usepackage{natbib}
        \bibpunct[, ]{(}{)}{;}{a}{}{,}
\begin{document}
	   \title{Interferometric properties of pulsating C-rich AGB stars}
	\subtitle{I. Intensity profiles and uniform disc diameters of dynamic model atmospheres}

	   \author{C. Paladini\inst{1}\and B. Aringer\inst{1,2} \and J. Hron\inst{1} \and
             W. Nowotny\inst{1} \and S. Sacuto\inst{1} \and S. H\"ofner\inst{3}
           }

	   \institute{Department of Astronomy, University of Vienna,
	              T\"urkenschanzstrasse 17, A-1180 Wien, Austria\\
	              \email{claudia.paladini@univie.ac.at}
	         \and
                 INAF-OAPD, Vicolo dell'Osservatorio 5, 35122 Padova, Italy\\
                 \and
	              Department of Physics and Astronomy, 
                      Uppsala University, Box 515, SE-75120, Uppsala, Sweden\\
	             }

	   \date{Received ; Accepted }

	 
	  \abstract
	  {} 
	   {We present the first theoretical study on center-to-limb variation (CLV) 
             properties and relative radius interpretation for narrow 
             and broad-band filters, on the basis of a set of 
             dynamic model atmospheres of \mbox{C-rich} AGB stars. We computed visibility profiles and the 
             equivalent uniform disc radii
             (UD-radii) in order to investigate the dependence of these quantities upon the wavelength and pulsation phase.}
	    {After an accurate morphological analysis of the visibility and intensity profiles
             determined in narrow and broad-band filter,
             we fitted our visibility profiles with a UD function simulating the observational approach. UD-radii have been 
             computed using three different fitting-methods to investigate the influence of the sampling of the visibility profile: 
             single point, two points and least square method.}
	   {The intensity and visibility profiles of models characterized by \mbox{mass loss} show a behaviour very different from a UD. 
             We found that UD-radii are wavelength dependent and this dependence 
             is stronger if \mbox{mass loss} is present. Strong opacity contributions from C$_2$H$_2$
           affect all radius measurements at 3 $\mu$m and in the N-band, resulting in higher values for the UD-radii. 
           The predicted behaviour of UD-radii versus phase is complicated in the case of models with \mbox{mass loss}, 
           while the radial changes are almost sinusoidal for the models without \mbox{mass loss}. 
           Compared to the \mbox{M-type} stars, for the C-stars no windows
           for measuring the pure continuum are available.}
	   {}


	   \maketitle
	%

	\section{Introduction}

	The Asymptotic Giant Branch (AGB) is a late evolutionary stage of stars with masses less than about $8 \, M_{\odot}$.
	The objects on the AGB are characterised by a degenerate C-O core
	and He/H-burning shells, a convective envelope and a very extended atmosphere containing molecules and
	in many cases even dust grains. The atmospheres are affected by the pulsation of the interior 
        creating shocks in the outer layers. 
	Due to the third dredge-up the AGB stars may have C/O\,$>1$ \citep{ibe83} and their spectra are 
	dominated by features of carbon species like C$_2$, C$_2$H$_2$, C$_3$,
	CN, HCN \citep{goe81, joy98, lan00, loi01, yam00}. These are the classical Carbon stars. 
        In such objects dust is mainly present as amorphous carbon grains.
	
	Studying stellar atmospheres is essential for a better comprehension of this late stage of stellar evolution,
	to understand the complicated interaction of pulsation and atmospheric structure, as well as 
        the processes of dust formation and mass loss.
        The atmospheres of C-rich AGB stars with no pronounced pulsation can be described by 
        hydrostatic models \citep{jor00, loi01}.
        As the stars evolve the effective temperatures decrease and the atmospheres become more extended. 
        At the same time the effects of time-dependent phenomena (dynamic processes) become important and static models are not
        a good approximation anymore. More sophisticated tools are needed for describing such objects
        i.e. dynamic model atmospheres. The status of modelling atmospheres of cool AGB stars is summarised in different reviews
        by \citet{wil00, woi03, hoe05, hoe07, hoe08}. Comparison of dynamic models with spectroscopy for C-rich stars 
        can be found in \citet[][henceforth GL04]{hro98, loi01, gau04}; 
        \citet{now05a, now05b, now05c, now09}. 

        Because of their large atmospheric extension and their high brightness in the red and infrared, AGB stars are
	perfect candidates for interferometric investigations.         
        Moreover, since most observed AGB stars are very far away (at distance larger than 100 pc), only the
        resolution reached by optical interferometry allows a study of the close circumstellar environment of these objects.
        
        The complex nature of their atmospheres opens 
        the question of how to 
        define a unique radius for these stars. 
        In the literature 
        several radius definitions were used, they are summarized in the reviews of \citet{bas91} and \citet[][henceforth S03]{sch03}.
        This is not a big problem when we deal with objects that have a less extended and almost hydrostatic atmosphere, 
        because in most cases the different definitions 
        of radius converge to a similar final result. However, one can find large differences
        when the envelope of the star is affected by the phenomena of pulsation and \mbox{mass loss}.
        In \citet{bas91} the authors suggest the use of the intensity radius that is ``a monochromatic (or filter-integrated) 
        quantity depending on wavelength or on the position
        and width of a specific filter''. This radius is determined from the center-to-limb variation (CLV) as the
        point of inflection of the curve, in other words as the point where the second derivative of the intensity profile
        is equal to zero. But to determine this value, a very accurate knowledge of the CLV has to be available.
        For AGB stars this is not always easy, since in 
        the case of an evolved object the intensity profile can present a quite complicated behaviour and only few points are measured. 
        A standard procedure in the case of a poorly measured CLV is to use analytical functions like the uniform disc (UD),
        a fully darkened disc (FDD) or a gaussian to fit to the profile. The result of this fit is the so-called fitting radius
        (S03).
        
        Currently, some theoretical and observational studies about the different properties 
        of the CLV and radius interpretations for \mbox{M-type} stars exist in the literature. They are summarised in the review of S03. 
        According to this paper, the CLV shapes can be classified in:
        
        \begin{itemize}
        \item \emph{small to moderate limb-darkening} when the behaviour of the profile is well fitted by a UD; 
        \item the \emph{gaussian-type CLV} is typical for extended atmospheres where 
        there is a big difference in temperature between the cool upper atmosphere layers and the deep layers where the continuum is formed; 
        \item the \emph{CLV with tail or protrusion-type extension} consists of two components,
        with a central part describable by the CLV of a near-continuum layer, and a tail shape given by the CLV of an outer shell;
        \item the \emph{uncommon CLVs} are profiles that show strange behaviours and are characteristic for complex extended 
          and cool atmospheres.
        \end{itemize}
        
        The fitting radius is usually converted to a monochromatic optical depth radius $R_{\lambda}$, 
        defined as the distance between the centre of the object and the layer with
        $\tau_{\lambda} = 1$ at a given wavelength. 
        But this conversion has to be done very carefully: even in the case of near-continuum 
        bandpasses, spectral features can contaminate the measurement of the fitting radius as shown by \citet{jac02} or \citet{ari08}, 
        and this will be not anymore a good approximation to the radius of the continuum layer.
        
        A filter radius, following the definition given by \citet{sch87} 
        \begin{equation}
        \centering
        R_{filt} = \frac{\int R_{\lambda}I_{\lambda}^{c}f_{\lambda}d\lambda}{\int I_{\lambda}^{c}f_{\lambda}d\lambda}
        \label{rfil}
        \end{equation}
        requires a good knowledge of the filter transmission curve and the filter itself has to be chosen in a careful way
        to avoid impure-filter-like effects (see Sect.5 of S03 for more details).
        A very common definition of radius is the Rosseland radius that corresponds to the distance between
        the stellar centre and the layer with Rosseland otical depth $\tau_{\rm{Ross}} = 1$. As in the case of the monochromatic optical depth radius
        this quantity is not an observable, it is model dependent, molecular contaminations are the main effect that has to be taken in account. 
        
        A suitable window at 1.04 $\mu$m
        for determining the continuum radius of \mbox{M-type} stars was discussed by \citet{sch87}, \citet{hof98}, \citet{jac02} and  
        \citet{tej03a, tej03b}.
        The authors defined a narrow-bandpass located in this part of the spectrum, mostly free from contaminations of molecules
        and lines. They indicate this radius
        as the best choice for studying the geometric pulsation.
        Unfortunately such a window does not exist 
        in the case of \mbox{C-rich} AGB stars (cf. Sect.~\ref{ud.sect}), and some model-assumptions have to be
        made for defining a continuum radius. On the other hand, available dynamic atmospheric models for Carbon-stars \citep{hoe03}
	are more advanced than for \mbox{M-type} stars, the main reason being a better understanding of the dust formation.

 	Following the same approach as done for M-stars by Scholz and collaborators, we started
        an investigation on the CLV characteristics
	for a set of dynamic models of pulsating C-stars \citep{hro08}.
 
        In this work we present an overview on the main properties of the intensity and visibility
        profiles in narrow and broad-band filters for dynamic model atmospheres with and without \mbox{mass loss} (Sect.~\ref{prof.sect}). 
        We use then the UD function to determine the UD-radius
        of the models, and investigate the dependence of the latter upon wavelength and pulsation phase.
        We discuss also a new definition for a continuum radius for the carbon stars (Sect.~\ref{ud.sect}).   

	\section{Dynamic models and interferometry}
        \label{DMA.sect}
        
        \subsection{Overview of the dynamic models}
        The dynamic models used in this work are described in detail in \citet{hoe03}, 
        GL04 and \citet{now05a}. 
        
        The models are spherically symmetric and characterised by time-dependent dynamics and frequency-dependent radiative transfer.
        The radiative energy transport in the atmosphere and wind
        is computed using opacity sampling in typically
        50-60 wavelength points, assuming LTE, for both gas and dust.
        For the gas component, an energy equation is solved together
        with the dynamics, defining the gas temperature stratification
        (and allowing, in principle, for deviations from
        radiative equilibrium, e.g., in strong shocks).
        For the dust, on the other hand, the corresponding grain
        temperature is derived by assuming radiative equilibrium.
        Dust formation (in this sample only amorphous carbon dust is considered) is treated by the "moment method'' \citep{gai88, gau90}, while 
        the stellar pulsation is simulated by a piston at the inner boundary.
        Assuming a constant flux there this leads to a sinusoidal bolometric 
        lightcurve which can be used to assign bolometric phases to each 'snapshot' 
        of the model with $\phi$\,=\,0 corresponding to phases of maximum light.

        In Table~\ref{mod.tbl} the main parameters of the models used here are summarised. 
        Luminosity $L_\star$, effective temperature $T_\star$ and C/O ratio refer to the initial hydrostatic model
        used to start the dynamic computation.
        This model can be compared with classic model atmospheres like MARCS \citep{gus08}. The stellar radius $R_{\star}$ 
        of the initial hydrostatic model is obtained from $L_{\star}$ and $T_{\star}$ and it corresponds to the Rosseland radius. 
        The piston velocity amplitude $\Delta u_{\rm{p}}$ and the period $P$ are also input parameters and rule the pulsation, while 
        {the \mbox{mass loss} rate} $\dot{\rm{M}}$ and the mean degree of condensation $\langle f_{\rm{c}}\rangle$ 
        are a result of the dynamic computations. 
        The free parameter $f_{\rm{L}}$ is 
        introduced to adjust the amplitude of the luminosity variation at the inner boundary in a way that 
        $L_{\rm in}(t)\propto f_{\rm{L}}\cdot R^{2}_{\rm{in}}(t)$ (GL04).
 
        \begin{table*}[ht!]
	      \caption{Parameters of the dynamic model atmospheres used in this work (see also GL04). Luminosity ($L_*$), 
                temperature ($T_*$), Rosseland radius ($R_{\star}$) and C/O refered to the 
              initial hydrostatic model, amplitude of the piston velocity ($\Delta u_{\rm{p}}$), period (P) and $f_{\rm{L}}$ are 
              input parameters for the dynamic calculations, 
              while {\mbox{mass loss} rate} ($\dot{\rm{M}}$) and the mean degree of condensation $\langle f_{\rm{c}} \rangle$ are a result.
              All the initial models have a mass $M_{\star} = 1\, M_{\odot}$ and solar metallicity. 
              Models in bold are the reference chosen for this 
              work. See also text (Sect.~\ref{DMA.sect}).}
	         \label{mod.tbl}
                 \centering
	         \begin{tabular}{l l l l l l l l l l l}
	            \hline
	            \hline
	            \noalign{\smallskip}
	            Model      &  $L_\star$ & $T_{\star}$ & $R_{\star}$ & C/O & $\Delta u_{\rm{p}}$ & $P$ &$f_{\rm{L}}$& $\Delta\rm{M}_{bol}$ & $\langle f_{\rm{c}}\rangle$ &$\dot{M}$\\
                               & [$L_\odot$]&     [K]    &[$R_{\odot}$]&     & [km\,s$^{-1}$]      & [d]&           & [mag] &&[M$_\odot$ yr$^{-1}$]\\
                    \hline
                    \textbf{D1} & \textbf{10\,000}    &\textbf{2600}&492& \textbf{1.4} & \textbf{4.0}& \textbf{490} & \textbf{2.0} & \textbf{0.86}&\textbf{0.28} & $\mathbf{4.3 \cdot 10^{-6}}$\\  
                     D2 & 10\,000    &2600&492& 1.4 & 4.0 & 525 & 1.0& 0.42  & 0.37& $5.9\cdot 10^{-6}$ \\  
                     D3 &  7000      &2600&412& 1.4 & 6.0 & 490 & 1.5& 1.07  & 0.40& $2.5\cdot 10^{-6}$ \\  
                     D4 &  7000      &2800&355& 1.4 & 4.0 & 390 & 1.0& 0.42  & 0.28& $2.4\cdot 10^{-6}$ \\  
                     D5 &  7000      &2800&355& 1.4 & 5.0 & 390 & 1.0& 0.53  & 0.33& $3.5\cdot 10^{-6}$ \\  
                     \hline
                     \textbf{N1} &  \textbf{5200}   &\textbf{3200 }&\textbf{234}&\textbf{1.1} &\textbf{2.0}&\textbf{295} & \textbf{1.0} & \textbf{0.13}&\textbf{--}&\textbf{--}\\  
                     N2 &  7000      &3000&309& 1.1 & 2.0 &390&1.0               & 0.23 &--& -- \\  
                     N3 &  7000      &3000&309& 1.4 & 2.0 &390&1.0               & 0.24 &--& -- \\  
                     N4 &  7000      &3000&309& 1.4 & 4.0 &390&1.0               & 0.48 &--& -- \\  
                     N5 &  7000      &2800&355& 1.1 & 4.0 &390&1.0               & 0.42 &--& -- \\  
 	            \hline
	        
	\end{tabular}
	   \end{table*}

        The models used in this work were chosen with the purpose to investigate the effects of 
        the different parameters on radius measurements: mainly \mbox{mass loss}
        but also effective temperature, C/O and piston amplitude. The sample can be divided into
        two main groups, keeping this purpose in mind: models with and without \mbox{mass loss} 
        (in Table~\ref{mod.tbl} series D and N, respectively). 
    
\begin{figure}
	   \centering
	   \includegraphics[width = 8.8cm]{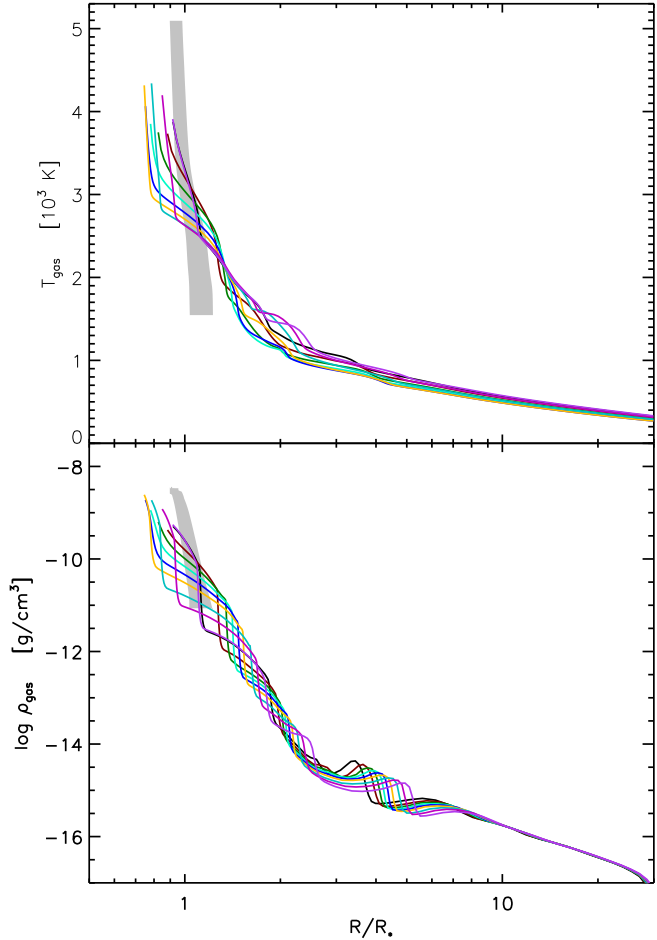}
 	   \caption{Radial structure of the dynamic model D1. Gas temperature (upper panel) and density (lower panel) 
             are plotted for different phases during the pulsation period. 
             For comparison the grey-shaded area marks the range where the structure of the model N1 at different phases can be found.
             More details about the structure of the models can be found in the text and in \citet{now05a, now05b}.}
	   \label{struct.fig}%
	    \end{figure}

        It was already mentioned that the pulsation of the inner
        layers can influence the outer atmosphere of the star. The propagation of the shock waves causes a levitation of the
        outer layers and in certain cases formation of dust grains can take place. The radiation pressure on the dust grains
        may drive a stellar wind characterized by terminal velocities $V_\infty$ and \mbox{mass loss} rates $\dot{\rm{M}}$. 
        In Fig.~\ref{struct.fig} an example of the radial structure of the dynamic model 
        atmosphere D1 (scaled to $R_\star$) is plotted for different phases.
        For comparison, the area covered by the model N1 is also shown grey-shaded.  
        In particular
        we show the effect of the dust-driven wind that appears on the model structure D1. 
        The different atmospheric extensions of the two models are easily noted.

        In the series D of Table~\ref{mod.tbl} two models, D1 and D2, differ from the others by a higher luminosity ($L_\star$).
        The only distinction between these two models are different values
        of the period $P$ and the parameter $f_{\rm{L}}$.
        The effect on the structure of the atmosphere is that dust shells observed at few $R/R_\star$ are more pronounced in the case of D2.   
        The models D4 and D5 have the same luminosity, temperature and period but different  
        piston velocities and different resulting degrees of dust condensation.
        In the structure of D4, compared with D5, there is a smoother transition from the region 
        dominated by pulsation to the one dominated by
        the dust-driven wind \citep[see Fig.~2 of][]{now05a}.
        
        In the N set of models we can distinguish N1 by its higher temperature and lower luminosity.
        All the other models have the same luminosity but differ by other parameters.
        N3 and N4 are an example for studying the effect of different piston velocities on models without \mbox{mass loss}.
        The atmospheric stucture of N4 is more extended but the outer layers are not yet cool and dense enough
        for dust formation.
        N2 and N3 have a different C/O ratio that causes a difference in the spectral features but also 
        in the estimation of the diameter at certain wavelengths (Sect.~\ref{ud.sect}).
        
        We will use model D1 as reference among the ones with \mbox{mass loss}. 
        This model was compared in GL04 with observed
        spectra of the star S\,Cep, it was shown that different 
        phases of D1 can reproduce the S Cep spectra shortward of $\approx\,4$\,$\mu$m. 
        The same model is in qualitative agreement with observed CO and CN line profiles variations of S\,Cep and other
        Miras \citep{now05a, now05b}.
        Concerning models without \mbox{mass loss} we use model N1 as a reference, which gives 
        the best fit of spectra and colour measurements for TX\,Psc in GL04.
        Each one of the reference models represents also the most extreme case in its sub-group.
        
	\subsection{Synthetic intensity and visibility profiles}
        \label{prof.sect}
                
        To compute intensity and visibility profiles on the basis of the models we proceeded in the following way.

        For a given atmospheric structure (where temperature and density
        stratifications both are taken directly from the dynamical simulation)
        opacities were computed
        using the COMA code, originally developed by \citet{ari00}: 
        all important species, i.e. CO, CH, C$_2$, C$_3$, C$_2$H$_2$, HCN and CN, are included.
        The data sources used for the continuous opacity 
        are the ones presented in \citet{led09}. Voigt profiles are used for atomic lines,
        while Doppler profiles are computed for the molecules. Conditions of LTE and a microturbulence value of $\xi = 2.5\,\rm{km\,s}^{-1}$
        are assumed.
        All references to the molecular data are listed in Table~2 of \citet{led09}. The source for the
        amorphous carbon opacity is \citet{rou91}.
        These input data are in agreement with the
        previous works \citep{now05c, led09, ari09}, and with the opacities used for the computation of the dynamic models.
        Information 
        on recent updates on the COMA code and opacity data can be found in \citet{now05c}, \citet{led09}, \citet{ari09} and \citet{now09}.
        The resolution of the opacity sampling adopted in our computations is 18\,000. 
        
        The resulting opacities are used as input for a spherical radiative 
        transfer code that computes the spectrum at a selected resolution and 
        for a chosen wavelength range. In addition, the code computes spatial 
        intensity profiles for every frequency point of the calculation.
        Velocity effects were not taken into account in the radiative transfer.

        The monochromatic spatial intensity profiles were subsequently convolved with different filter curves 
        (narrow and broad-band) in the near- to mid-IR range to obtain averaged intensity profiles.
        The narrow-band filters are rectangular (transmission $1$), while the broad-band filters are the standard filters 
        from \citet{bes88}.
        The visibility profiles are the two-dimensional Fourier transform 
        of the intensity distribution of the object, but under the condition 
        of spherical symmetry (as for our 1D models), this reduces to the
        more simple Hankel transform of the intensity  profiles \citep[e.g.][]{bra65}.
       
        The intensity and visibility profiles in the case of the broad-band filters were computed with the following procedure.
        Each broad-band was split into a set of 10 narrow-band subfilters\footnote{In principle this should be done monochromatically
        but  for computational reasons the usual number 
        of narrow-band subfilters was limited to 10.
        However, we compared the resulting broad-band filters computed
        with a set of 10 narrow-band subfilters with the one computed with a set of 100 and 200 subfilters without finding any significant
        difference in the resulting broad-band visibility profile. 
        We want to stress also that the maximum resolution for the computation of the broad-band filter 
        is given by the opacity-sampling resolution.} and the squared broad-band visibility was computed as 
 
        \begin{equation}
          V_{\mathrm{broad}}^2 = 
          \frac{\sum\limits_{{i}}(S_{{i}}^2 F_{{i}}^2 V_{{i}}^2)}
               {\sum\limits_{{i}}(S_{{i}}^2 F_{{i}}^2 )}\,,
          \label{broad.formula}       
        \end{equation}

        \noindent where the sum is calculated over all the narrow subfilters. $S_{\rm{i}}$ is the transfer function of the broad-band filter 
        within the corresponding narrow-band subfilter, 
        $F_i$ is the flux integrated over the corresponding narrow-band subfilter, and $V_i$ is the visibility of the
        narrow-band subfilter computed from the average intensity profile for the corresponding filters.
        
        In the following two sections we will discuss the characteristics of the intensity and visibility profiles for different models 
        and the effects when using different filters. In the case of broad-band filters also the so called 
        \emph{band width smearing effect} \citep{ver05,ker03} will be discussed.

%
	   \begin{table}
	      \caption[]{Definitions of the narrow and broad band filters. Listed are the names 
                of the bands, characteristic features (for narrow-bands), 
                central wavelengths, bandwidths and the morphological classes (see Sect.~\ref{synth.sect}).
                In bold are indicated the filters chosen to represent the respective morphological class. The designation \emph{"cont"} 
                means that the spectrum includes no 
                strong molecular absorption, but this does not necessarily represent the real continuum.}

	         \label{filt.tbl}
	         \begin{tabular}{l l l l l}
	            \hline
	            \hline
                    \noalign{\smallskip}
	            Filter      &  Features & $\lambda_{\rm{c}}$ &  $\Delta \lambda$ & Class\\
	                        &           &   [$\mu$m]        & [$\mu$m]\\
	            \hline
	            1.01 & C$_2$, CN           &   1.01  & 0.02 & I \\ 
	            1.09 & CN                  &   1.09  & 0.02 & I\\ 
        	    1.11 & "cont"              &   1.11  & 0.02 & I\\ 
	            1.51 & CN+                 &   1.51  & 0.02 & I\\ 
	            \textbf{1.53} & C$_2$H$_2$, HCN, CN & 1.53  &0.02 & \textbf{I}\\ 
	            1.575 & CN                  &   1.575 & 0.05 & I\\
	            1.625 & CN, CO              &   1.625  & 0.05 & I\\		            
                    1.675 & "cont"              &   1.675 & 0.05 & I\\ 		            
                    1.775 & C$_2$               &   1.775 & 0.05 & I\\ 		            
                    2.175 & CN                  &   2.175 & 0.05 & I\\ 	            
	            2.375 & CO                  &   2.375 & 0.05 & I\\ 
	            \textbf{3.175} &C$_2$H$_2$, HCN, CN & 3.175 & 0.05 & \textbf{III} \\ 
	            3.525 & "cont"              &   3.525 & 0.05 & II\\            	            
	            3.775 & C$_2$H$_2$          &   3.775 & 0.05 & II\\ 
	            3.825 & C$_2$H$_2$          &   3.825 & 0.05 & II\\ 
	            3.975 & "cont"              &   3.975 & 0.05 & II\\ 
	            \textbf{9.975} & "cont" & 9.975 & 0.05 & \textbf{II}\\ 
	          11.025  & "cont"              &  11.025 & 0.05 & II\\ 	            	         
                  12.025  & C$_2$H$_2$          &  12.025 & 0.05 & III\\ 
	          12.475  & C$_2$H$_2$          &  12.475 & 0.05 & III\\ 	            
	          12.775  & C$_2$H$_2$          &  12.775 & 0.05 & III\\ 	
                  \hline
	            J    &                     &   1.26  & 0.44&\\ 
	            H    &                     &   1.65  & 0.38&\\ 
	            K    &                     &   2.21  & 0.54&\\ 
	            L$^\prime$  &               &   3.81  & 0.74&\\ 
	            \hline	        
	\end{tabular}
	   \end{table}

        \section{Synthetic profiles for narrow-band filters}
        \label{synth.sect}

\begin{figure*}
	   \centering
	   \includegraphics[width = 17.cm]{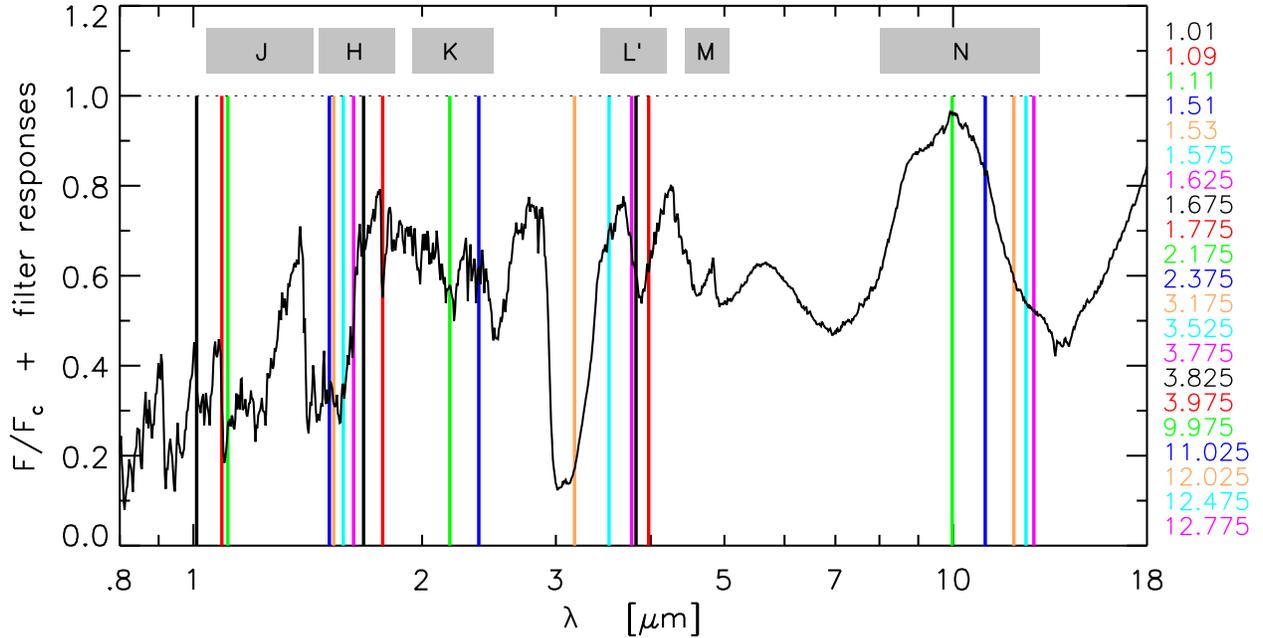}
 	   \caption{Synthetic spectrum ($R$\,=\,360) calculated on the basis of the hydrostatic intial atmospheric structure of model D1. 
             All necessary opacity sources (continuous, atoms, molecules, dust) are included for the spectral synthesis, (see Fig.~4 
             of \citet{ari09} for the individual contributions of the different molecular species). 
             The flux is normalised to a calculation for which only the continuous opacities were taken into account. 
             Overplotted are the narrow-band filters as defined in Table\,2, for which the central 
             wavelengths in $[\mu m]$ are given in the legend. In addition, the ranges of the broad-band filters are marked grey-shaded..}
           \label{filt.fig}%
	    \end{figure*}

        Aiming to study effects of different molecular features characteristic for C-star spectra and somewhat
        following the approach sketched in \citet{bes89,bes96}, 21 narrow-band filters  
        have been defined in the near-to mid-IR range with a ``resolution'' ranging from 50 to 200. 
        Figure~\ref{filt.fig} shows the location of the filters on top of a typical synthetic (continuum normalised) 
        spectrum of a C-type star for illustration. Table~\ref{filt.tbl} gives an overview of our complete set 
        of filters with central wavelengths, band widths, and related spectral features. 
        A few of these filters are labeled "cont" meaning that the contamination with molecular features 
        is comparably low in this spectral region. It may be possible at extremely high spectral resolutions to find 
        very narrow windows in the spectrum of a C-star where no molecular lines are located (e.g. Fig.~6 in Nowotny et al. 2005b). 
        However, in general and with resolutions available nowadays the observed spectra of evolved C-type stars in the visual 
        and IR are crowded with absorption features. Due to such a blending the continuum level $F$/$F_{\rm c}$=1 
        is never reached as in Fig.~\ref{filt.fig}. This effect is more and more pronounced with lower effective 
        temperatures and decreasing spectral resolutions \citep[cf.][]{now09}. 

	   \begin{figure*}
	   \centering
           \begin{tabular}{c}
 	   \includegraphics[width = 8.8cm]{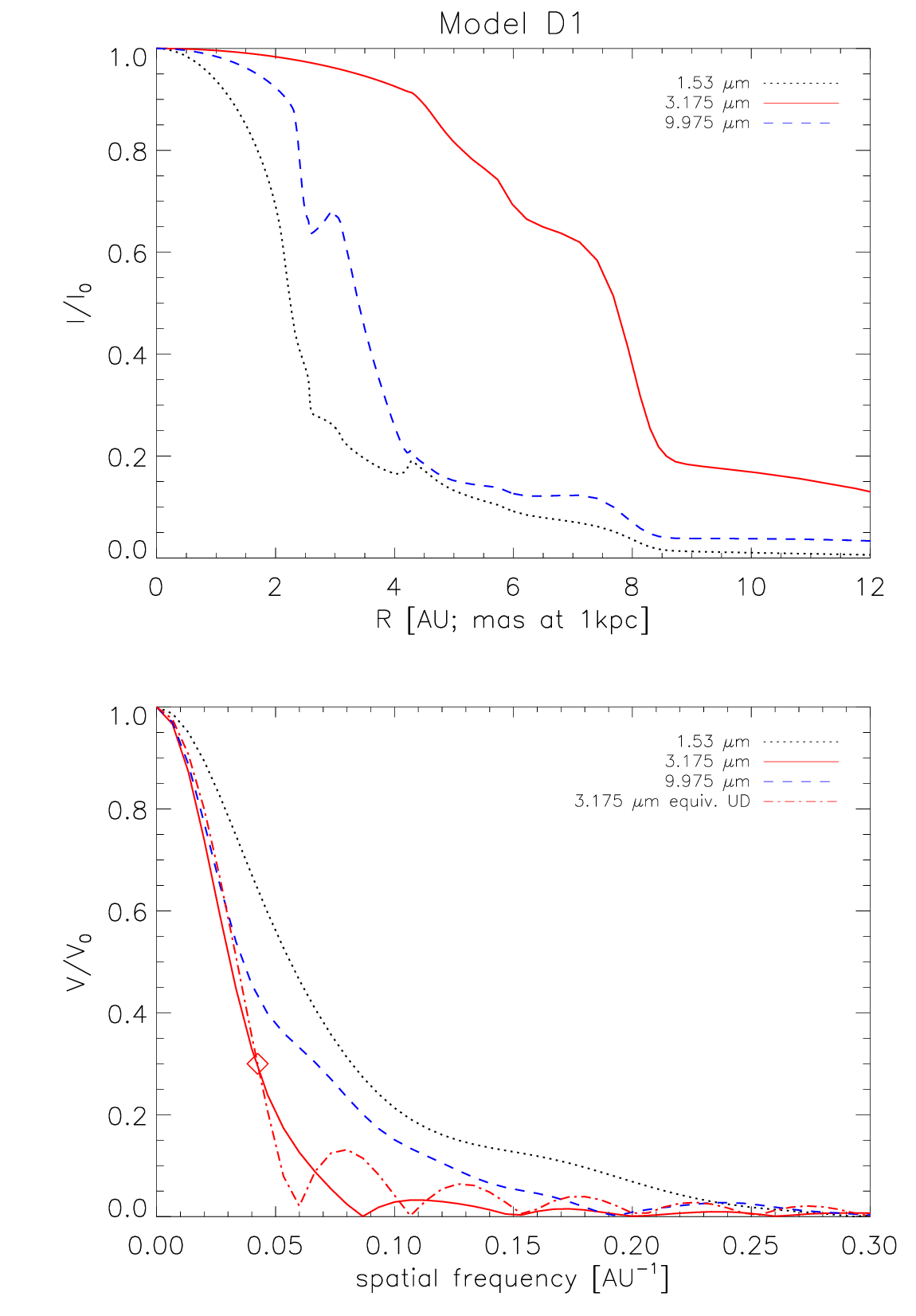}
          \includegraphics[width = 8.8cm]{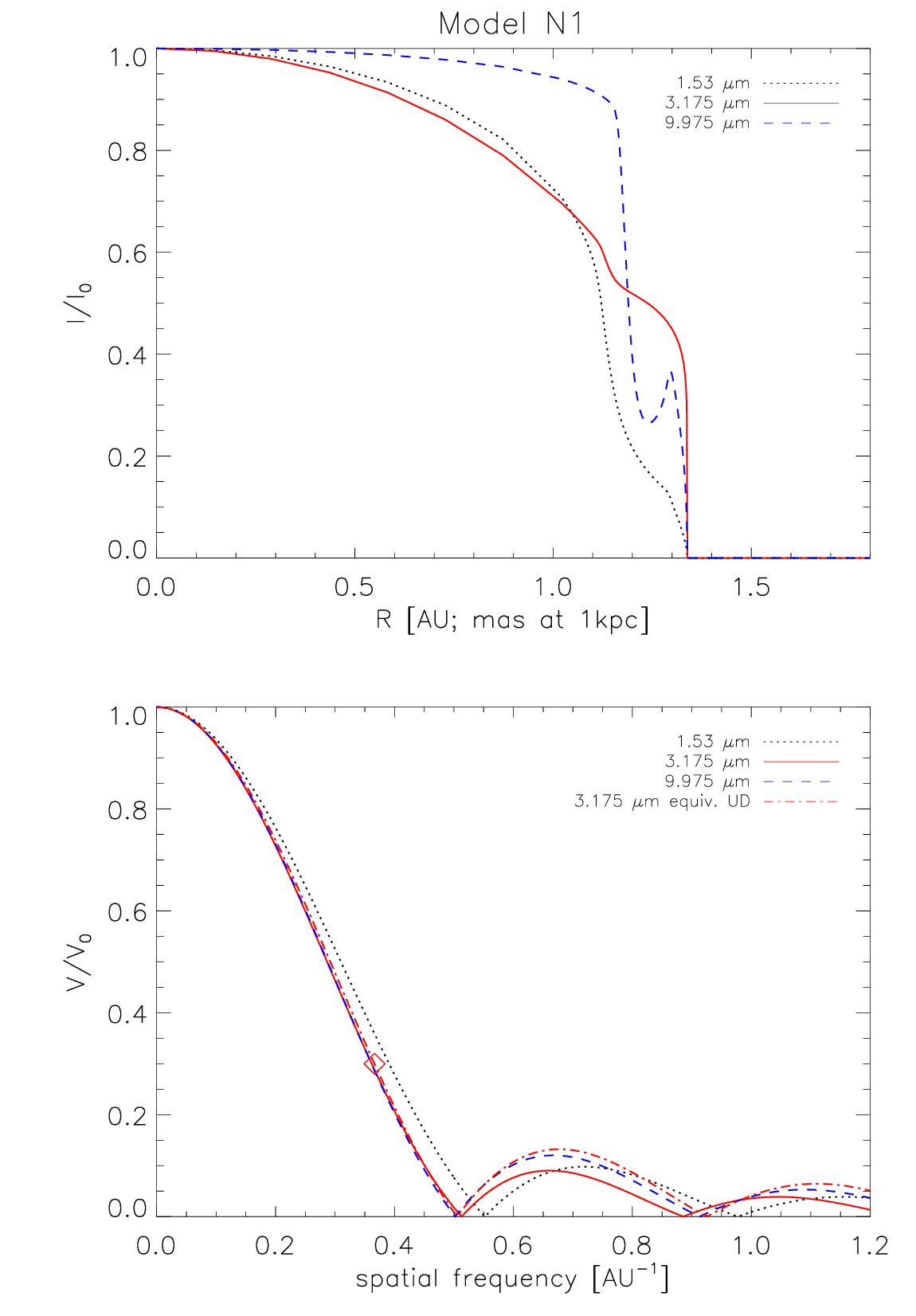} 
          \end{tabular}
	   \caption{Intensity and visibility profiles for the narrow-band filters at the maximum phase of 
             the mass-losing model D1 on the left side, and the model without \mbox{mass loss} N1
             on the right side. All the opacities are included in the computations. 
             Profiles for three representative filters are shown together with
           the equivalent UD function for the 3.175 $\mu$m filter.  
           The UD is fitted to the visibility point $V=0.3$ marked with the diamond (details in Sect.~\ref{ud.sect}).}
	   \label{prof.fig}%
	    \end{figure*}

        As a first step we study the morphology of different intensity and visibility profiles,
        then we proceed with an investigation of the main opacity contributors to the features that appear in the profiles. 
        For this purpose we compute for the two reference models 
        three different sets of intensity and visibility profiles: 
        (i) profiles including all necessary opacity sources i.e. continuous, atomic, molecular and dust opacities;  
        (ii) profiles where the opacities of dust were neglected, to analyse only molecular and atomic contributions, 
        and (iii) profiles for which only the continuous gas opacity
        is considered, to get theoretical continuum profiles.
        
        For the narrow-band filters, the intensity profiles of models with \mbox{mass loss} can be divided into three
        morphological classes listed in the last column of Table~\ref{filt.tbl}. It turns out that this classification
        is related mainly
        to the wavelength region with an exception for the filter $3.175\,\mu$m which belongs to the third class (that contains filters
        centered around $12\,\mu$m) instead of the second (that contains all the other filters in the $3\,\mu$m region plus the 9 and 11 $\mu$m
        filters).
        In Fig.~\ref{prof.fig} the intensity and visibility profiles for the representative filters of each class are illustrated.  
        Models with \mbox{mass loss} (left panels) 
        are more extended than models without \mbox{mass loss} (right), and their intensity profiles fit
        into the category of ``uncommon profiles'' of S03.

        The first class of profiles contains the filters 
        in the near IR from $1.01\,\mu$m to 
        $2.375\,\mu$m. 
        It can be well represented by filter 1.53 (C$_2$H$_2$, HCN, CN).
        Its morphology is characterised first by a limb darkened disc,
        followed by a bump due to molecular contributions, and a peak indicating the inner radius of the dust shell. 
        At 1.53 $\mu$m a feature due to 
        the molecular species of C$_2$H$_2$ and HCN
        is predicted by some of the cool dynamic models with wind (GL04). This feature 
        was observed in the spectra of cool \mbox{C-rich} Miras such as R\,Lep \citep{lan00}, 
        V\,Cyg, cya\,41 (faint Cygnus field star) and
        some other faint high Galactic latitude carbon stars \citep{joy98}. 
        
        The third class of profiles is represented by filter 3.175 (C$_2$H$_2$, HCN),  
        which is placed right in the center of an extremely strong
        absorption feature characteristic for C-star spectra (Fig.~\ref{filt.fig}). Observations in this spectral region
        are a very powerful tool for understanding the upper photosphere.
        Here, the inner central part of the intensity profile can be described by 
        an extended plateau due to strong molecular opacity. All the filters defined around 12 $\mu$m
        -- where C$_2$H$_2$ has an enhanced absorption --
        belong also to this class.
        
        The second class of profiles is intermediate between the two behaviours 
        illustrated before with an indication of an outer dust shell around $8\, AU$. 
        It is represented by filter 9.975 ("cont"), and it covers the behaviour of
        the filter at $11.025\, \mu$m plus those defined in the 3 $\mu$m region (except 3.175 $\mu$m). 

        In the Fourier space the three representative visibility profiles are also markedly different (lower left panel of Fig.~\ref{prof.fig}).
        The UD function, computed by fitting the point at
        visibility\footnote{All the visibility values $V$ have to be interpreted as normalized visibilities $V/V_0$.} $V= 0.3$ of the 3.175 $\mu$m 
        profile (for details see Sect.~\ref{ud.sect}), is shown for comparison
        in Fig.~\ref{prof.fig}. The visibility profiles from the model and the UD are different 
        at visibility $< 0.2$, in this region the model 
        profile is less steep and has a less pronounced second lobe compared with the UD.
                
        In the right panel of Fig. \ref{prof.fig} intensity and visibility profiles for the phase of maximum luminosity
        of the reference model without {\mbox{mass loss} are represented. As expected from the model structures, the intensity profiles
        are more compact than the ones for the models with \mbox{mass loss}. In the intensity profile corresponding to the filter 9.975, around $1.3\,AU$, 
        a peak due to the molecular opacity of C$_3$ is visible.
        \citet{jor00} showed that C$_3$ is particularly sensitive to the
        temperature-pressure structure and, hence, tends to be found in narrower regions than other molecules.
        The peak visible at 3 AU of the profile of the model D1 is due mainly to C$_3$ but
        also to other molecular and dust opacity contributions. 
        These $9.975~\mu$m peaks in the intensity profiles mean a marginally "ring''-type appearence of the disc.
        The shape of the intensity profiles for the models without \mbox{mass loss} can be classified 
        as having small to moderate limb-darkening (S03).
        In the Fourier space, the visibility profiles of the three filters for the series N look very similar except for tiny 
        differences especially in the second lobe. In all the cases they can be well approximated
        by a UD in the first lobe. 
        
        Plots of theoretical continuum profiles for both reference models are very similar to UDs.
        The classification of filters is valid also for all the other models of our sample listed in Table~\ref{mod.tbl}.

        %

        %
	   \begin{figure*}
	   \centering
           \begin{tabular}{c c}
	   \includegraphics[width = 8.8 cm]{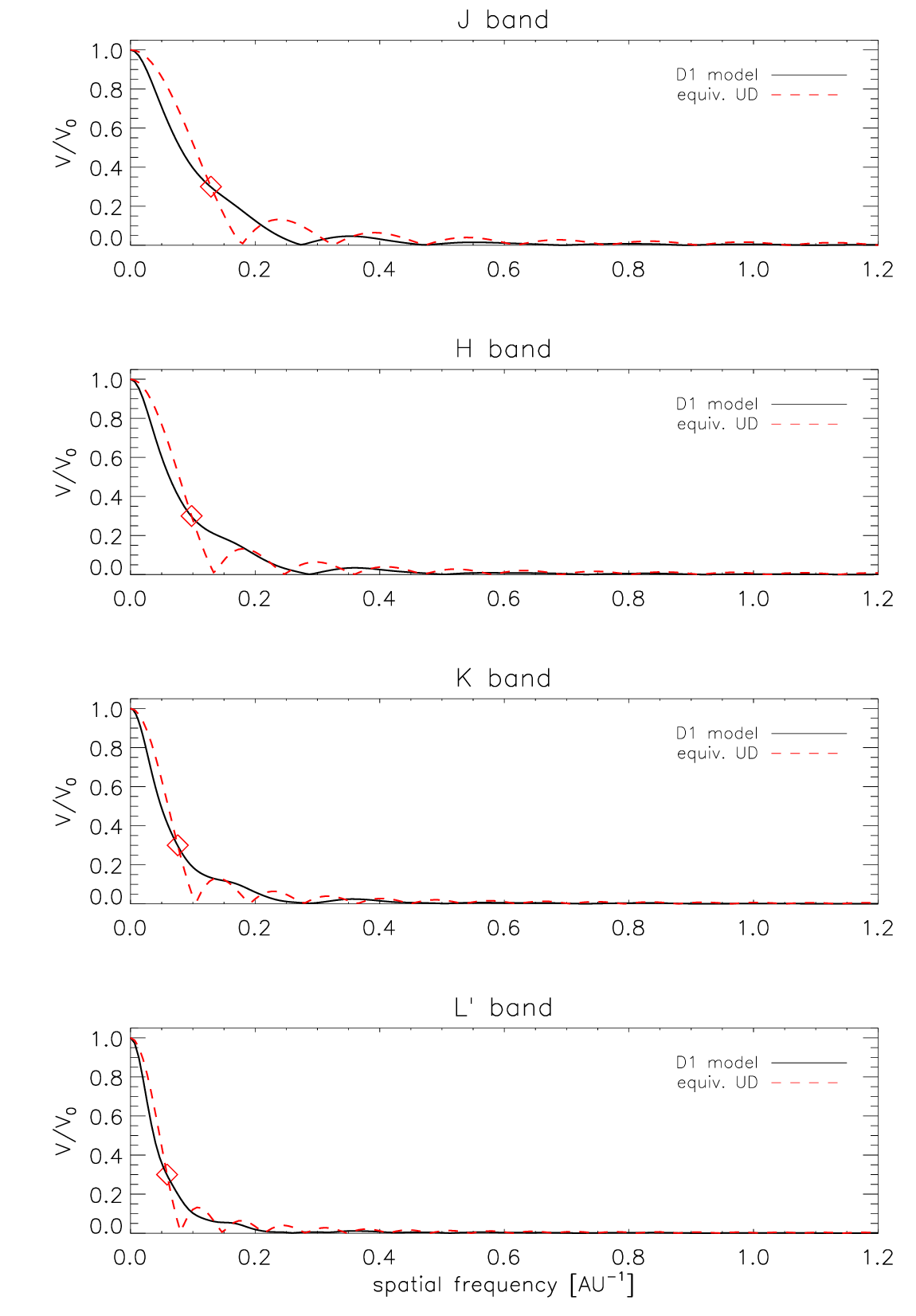}
	   \includegraphics[width = 8.8 cm]{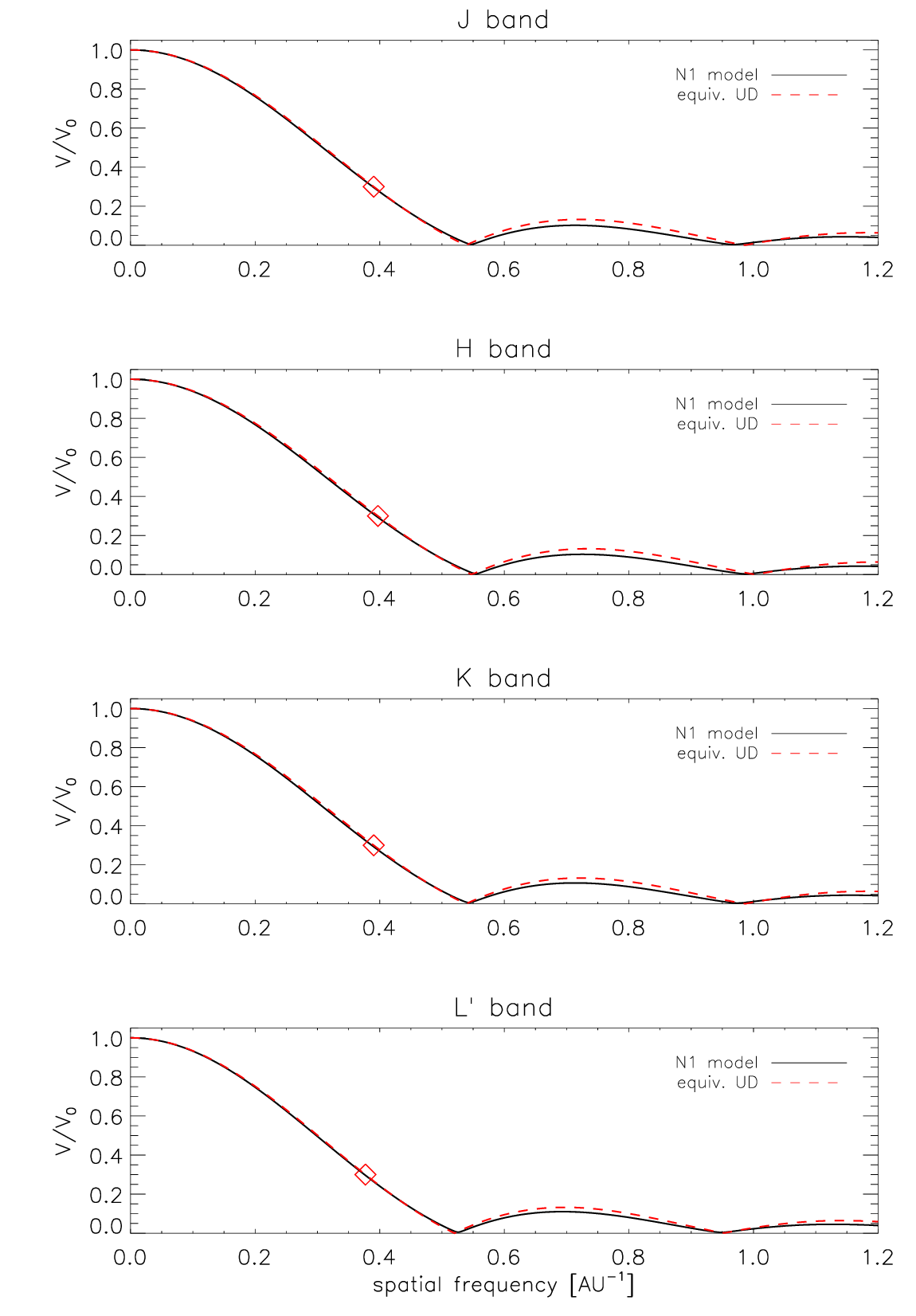}
           \end{tabular}
	   \caption{Visibility profiles computed in broad-band filters for the maximum phase of the model D1 with 
             \mbox{mass loss} on the left side, and the model N1 without \mbox{mass loss}
             on the right side.
             The solid line corresponds to the model while the dashed line is the equivalent UD 
             computed fitting one point at visibility $V = 0.3$.}
	   \label{broadprof.fig}%
	    \end{figure*}

        \section{Synthetic profiles for broad-band filters}
  
        We investigated also intensity and visibility profiles for
        the broad-band filters J, H, K, L$^\prime$, because several
        observations of C-stars exist in these filters.

        In Fig.~\ref{broadprof.fig} the resulting broad-band visibility profiles computed for the maximum 
        phase of the reference models are shown together with the corresponding UD visibility,
        which was calculated with the one-point-fit method at $V = 0.3$.
        In the case of model D1 (with \mbox{mass loss}) the broad-bands profiles show a behaviour completely different from the UD.        
        The slopes of the model profiles are always less steep than the corresponding UD at $V \lesssim 0.2$. 
        The synthetic visibility profile shows a characteristic tail-shape instead of the well pronounced second lobe that
        compare in the UD.
        The broad-band visibilities for the model N1 are well represented by a UD and, like in the case of the narrow-bands, 
        the major (but small) differences occur in the second lobe of the profiles.
        
        When a comparison to observations is performed in a broad-band filter particular attention has to be paid to the 
        smearing effect (also known as bandwidth smearing) which affects the measurements \citep{dav00,ker03,ver05}. 
        This is nothing more than the chromatic aberration due to the fact that observations are never monochromatic
        but integrated over a certain wavelength range defined by the broad-band filter curve. 
        Furthermore, each physical baseline $B$, corresponds to a different spatial frequency $u$ 
        with $u = \frac{B}{\lambda}$.
        A broad-band visibility, calculated at a given projected baseline, will result in a 
        superposition of all the monochromatic profiles over the width of the band.

        According to \citet{ker03}
         and \citet{wit04}, the bandwidth smearing is stronger at low visibilities
        close to the first zero of the visibility profile.
        We will demonstrate that also the shape of the profile plays an important role in the smearing effect. 
        
        We investigated this effect with the help of our reference models in the $J$, $H$, $K$ and $L^{\prime}$ filters.  
        As explained in Sect.~\ref{prof.sect} the broad-band profile was computed splitting the broad-band in 10 narrow-band subfilters
        and then averaging the resulting profiles. To take in account the bandwidth smearing, we split again the broad-band
        in the same set of 10 narrow-band subfilters.  
        The spatial frequencies are then converted from $AU^{-1}$ to baselines
        in meters, fixing the distance for an hypothetical object (500 pc) 
        and  using as corresponding ``monochromatic wavelength'' the respective central wavelength of the narrow-band filter. 
        Then the resulting broad-band profile was obtained using the above Eq.~(\ref{broad.formula}). 

        Figure~\ref{smearing-effect.fig} illustrates the bandwidth smearing effect for the \mbox{K band} for the
        two reference model atmospheres (D1 on the left and N1 on the right).
        It is clearly visible in the upper panels that the profiles of the 
        narrow-band subfilters reach the first zero at different baselines.
        The bandwidth smearing affects the broad-band profile in a way that it never reaches zero, as noted in
        \citet{ker03}. We compared the resulting visibility curve with a corresponding one  
        obtained without taking into account the bandwidth effect. The result of this experiment can be seen in the 
        lower panels of Fig.~\ref{smearing-effect.fig} where
        we plot the difference (in percentage) between the visibility with and without 
        taking into account the bandwidth smearing (BW and NBW respectively). 
        For both types of models -- with and without \mbox{mass loss} -- the smearing gives 
        a relative difference in the profiles that reaches at certain baselines more than $70\%$.
        For model N1 
        the major difference occurs near the first and the second zero and it is negligible elsewhere
        \citep[as already shown in][]{ker03,wit04}, while in the case of the model D1 the behaviour of the relative difference
        is less obvious.
        This result is confirmed also for the other broad-band profiles.
        However, it obviously depends on the shape of the visibility profile, and hence on the model.
        
        We conclude that in all the cases it is important to compute accurate
        profiles taking into account the bandwidth smearing. 

        %
	   \begin{figure*}
	   \centering
           \begin{tabular}{c c}	
	   \includegraphics[angle = 90, width = 8.8cm]{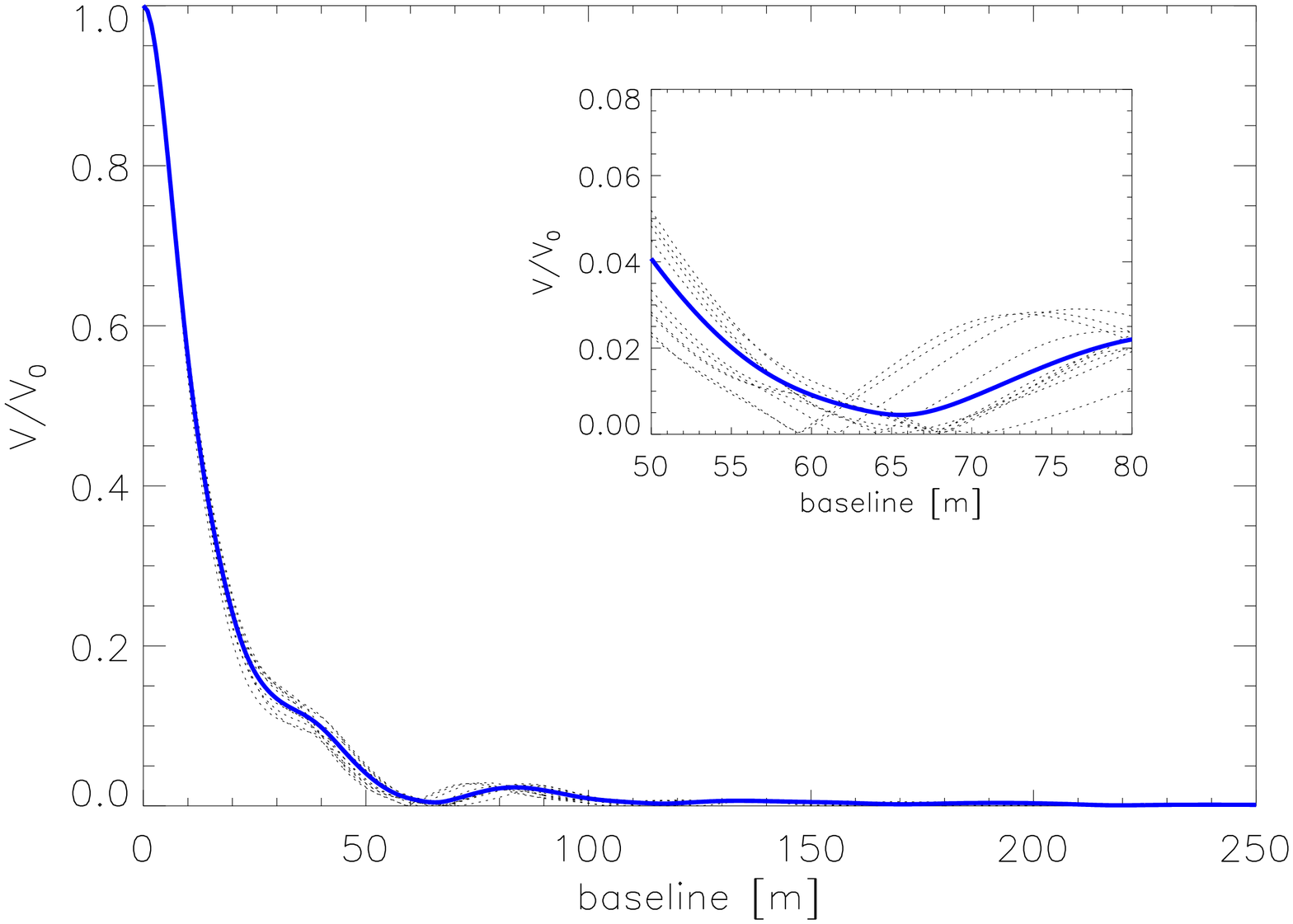}&
           \includegraphics[angle = 90, width = 8.8cm]{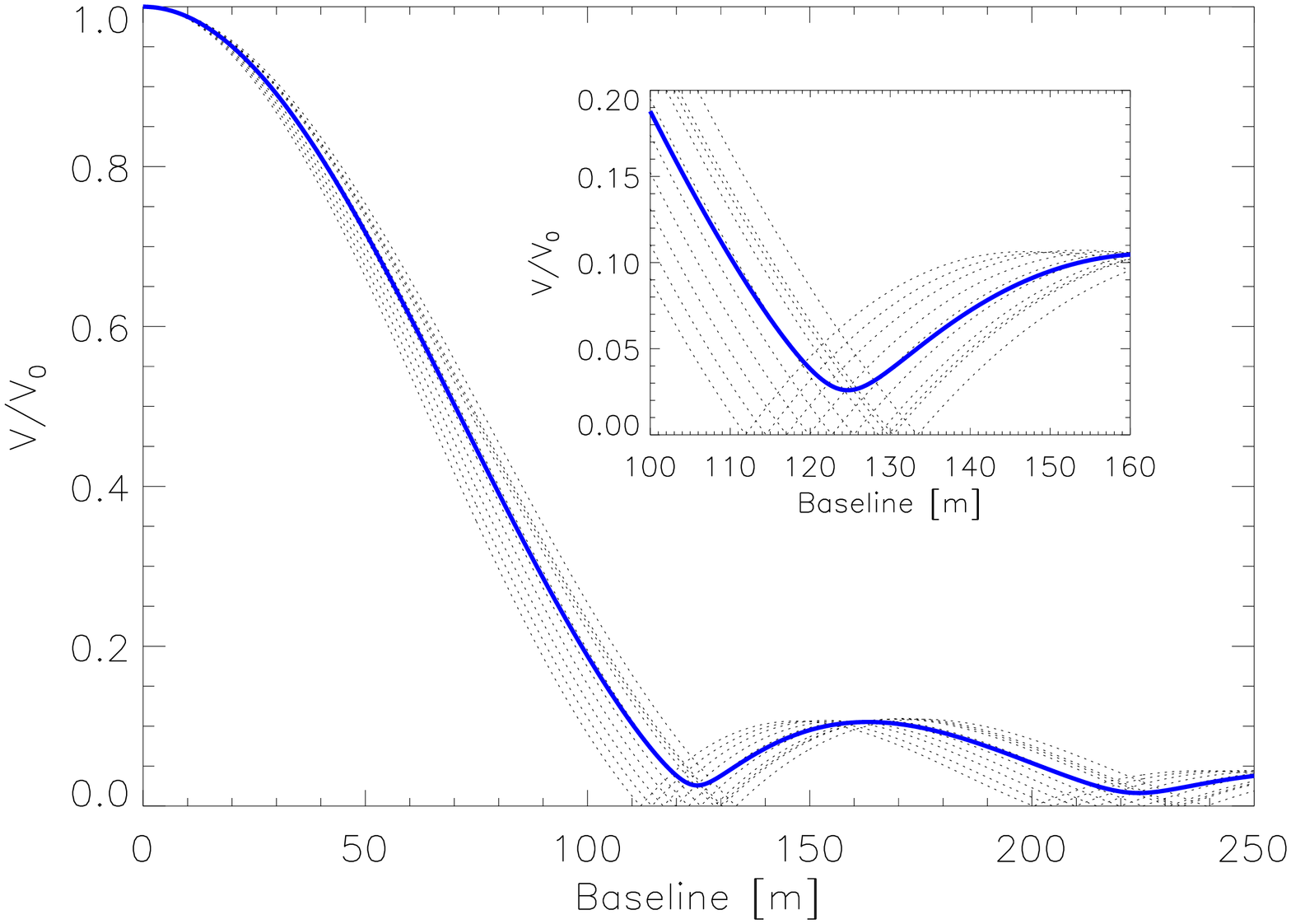} \\
           \includegraphics[angle = 90, width = 8.8cm]{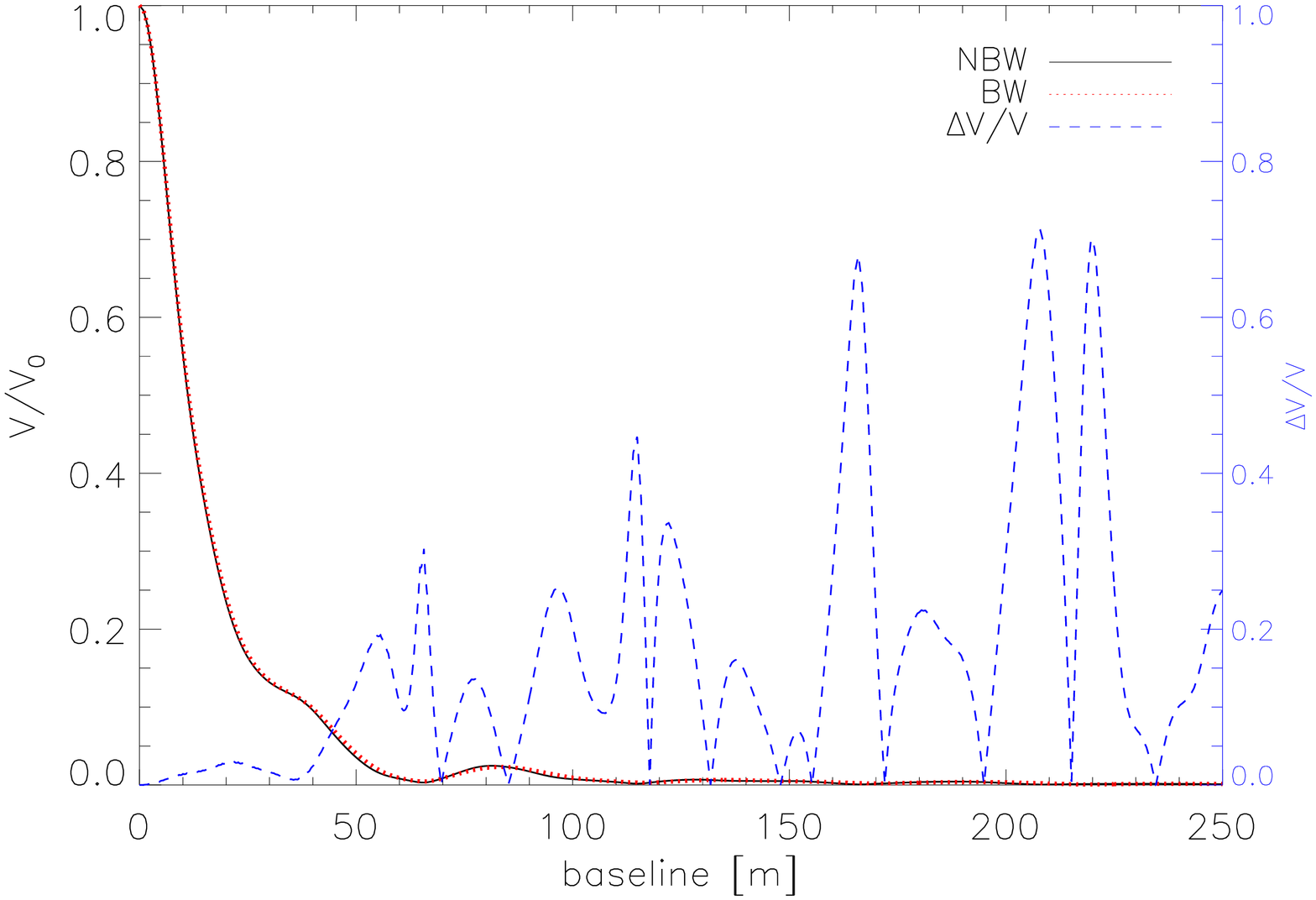}&
	   \includegraphics[angle = 90, width = 8.8cm]{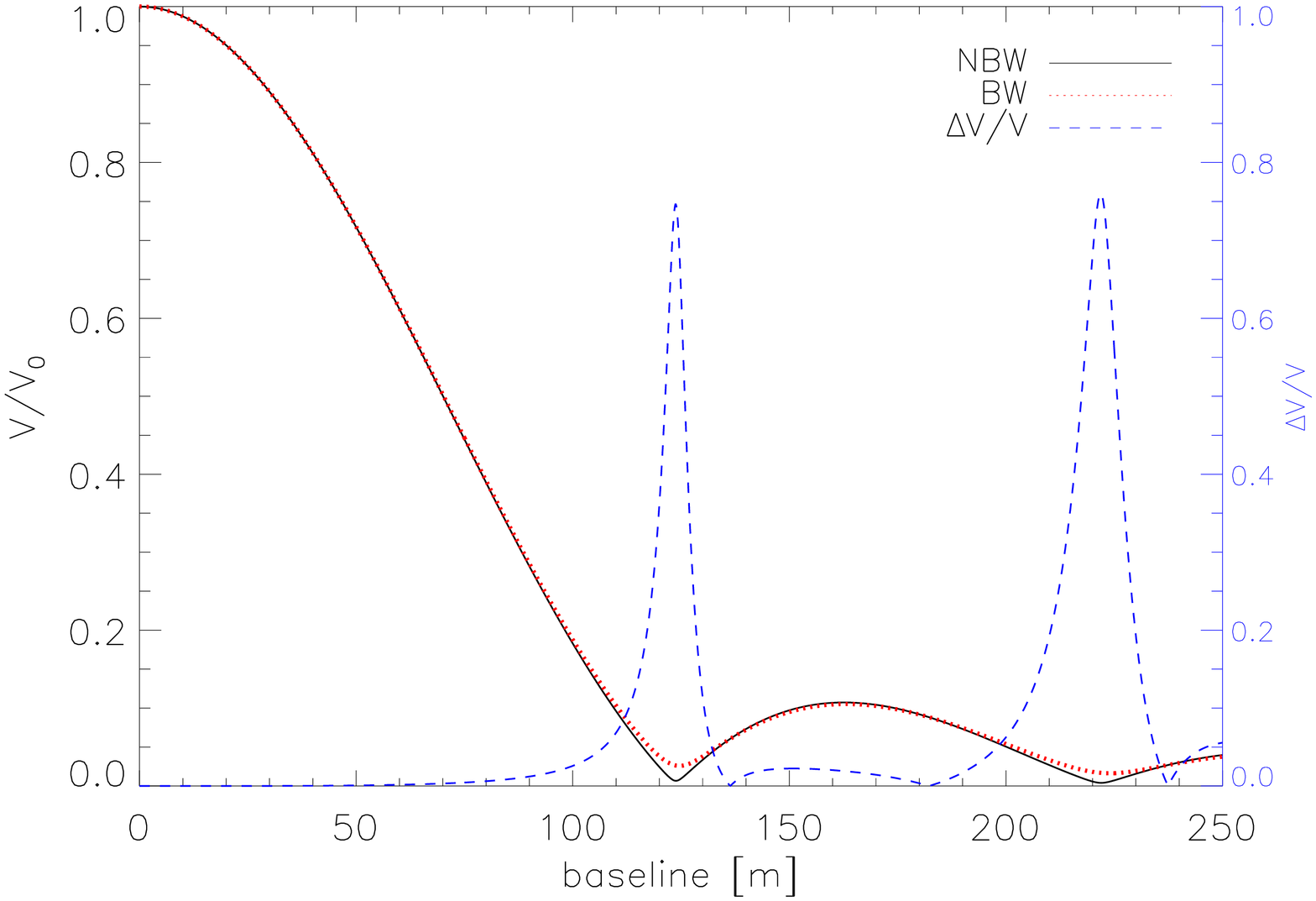}\\
           \end{tabular}
           \caption{Representation of the bandwidth smearing effect for the maximum of phase of the model D1 (left panels)
             and N1 (right panels) in the K band.
           In the upper panels visibility profiles as function of baseline for the individual narrow-band subfilters (dotted lines) 
           and for the resulting broad-band filter (solid line) -- computed
           taking into account the bandwidth smearing -- are plotted. In the inset, a zoom of the region around the first zero is represented.
           The lower panels show
           the visibility profiles computed with bandwidth smearing (BW, solid line), and profiles computed neglecting 
           the effect (NBW, dotted line). The dashed lines present the difference between 
           the two profiles (scale on the right y-axis).
           }
	   \label{smearing-effect.fig}%
	    \end{figure*}

           \section{Uniform disc (UD) diameters}
           \label{ud.sect}

           For all the models of the sample and all the considered filters 
           we computed the equivalent UD-radii using three different methods: 
           (i) a single point fit at visibility $V= 0.3$, (ii) a two-points fit at visibilities $V= 0.1$ and $0.4$, 
           and (iii) finally a non linear least square fit 
           using a Levenberg-Marquardt algorithm method, and taking into account all 
           the points with visibility $V > 0.1$.
           As already stressed in Sect.~5 of \citet{ire04a}, none of these procedures are
           fully consistent because the UD does not represent the real shape of the profile 
           (especially in the case of models with \mbox{mass loss}).
           However the UD is often chosen to interpret interferometric data with a small sampling of the visibility curve. 
           Thus, the three different methods are used in this work to investigate the influence of the sampling
           of the visibility profile, in particular if only a few visibility measurements
           are available.
           We use the visibility $V = 0.3$ for the one-point-fit following \citet{ire04a,ire04b} 
           who identified this point as a good estimator for the size of the object.
           While this is true for the \mbox{M-type} models presented in \citet{ire04a, ire04b},
           it is not obvious for our \mbox{C-rich} models with \mbox{mass loss}, which are very far from a UD 
           behaviour (Sect.~\ref{synth.sect}). 
           However keeping in mind that a UD is just a first approximation 
           we used this value also for our computations
           of the one-point fit.
           
           The aim of this section is to
           study the behaviour of the UD-radius versus wavelength and versus 
           phase, as well as the dependence on the
           different parameters of the models.
                      
           As we pointed out previously, while for the \mbox{M-type} stars the continuum 
           radius can be measured around 1.04 $\mu$m \citep[e.g.][]{jac02}, 
           our calculations show that the C-stars spectra do not provide any suitable window for this determination.
           However, the continuum radius is a relevant quantity for modeling of the stellar interior and pulsation.
           Therefore we computed for our sample of dynamic model atmospheres 
           a set of profiles taking into account only the continuous gas opacity.
           The resulting profiles correspond to the pure theoretical continuum absorption
           of the models. 
           An equivalent UD-continuum radius has been computed for 
           this theoretical "pure continuum" profiles. This allows us to study how the 
           UD-continuum radius depends upon wavelength and phase and 
           may give a possibility to convert observed radii to the continuum radius. 
         
       \subsection{Uniform disc radii as a function of wavelength}
         
	The UD-radii resulting from least square fits at the maximum phase 
        of all the models are plotted versus wavelength in Fig.~\ref{rlambda.fig}.
        They are scaled to $R_\star$ to minimize effects of the initial stellar parameters and enhance effects due to dynamics and dust formation.
        The UD-continuum  radii are represented with the cross symbols
        for D1 (upper panel) and N1 (lower panel).
        The major difference is found between models with and without mass loss. 
	In both cases the UD-radius generally increases with wavelength, but this 
	behaviour is more pronounced for mass-losing models.
	At 3.175 $\mu$m there is a notable "jump'' (value significantly higher than the surrounding) of the UD-radius due to the high C$_2$H$_2$ opacity.  
	The UD-continuum radius is mostly independent of wavelength and
        thus it can be adopted as a good reference radius (although not being observable for C-stars!).
        In fact, a small increase of its size can hardly be distinguished at longer wavelengths 
        (see Fig.~\ref{rlambda.fig}), and at $1.6\, \mu$m is visible the minimum opacity of H$^-$. 
        The ratio between the apparent size of the continuum radius at 11 and
        and 2.2 $\mu$m is 1.14 in the case of the model with \mbox{mass loss} D1 and 1.06 for the model without \mbox{mass loss}
        N1. This change in apparent size
        can be explained with the electron-hydrogen continuum, according to \citet{tat06}.
	All the models show a value of UD-continuum radius lower than the stellar radius $R_\star$. This is a consequence of the definition of $R_\star$.
         
        %
	   \begin{figure}[!ht]
	   \centering
	   \includegraphics[angle = 90, width = \hsize]{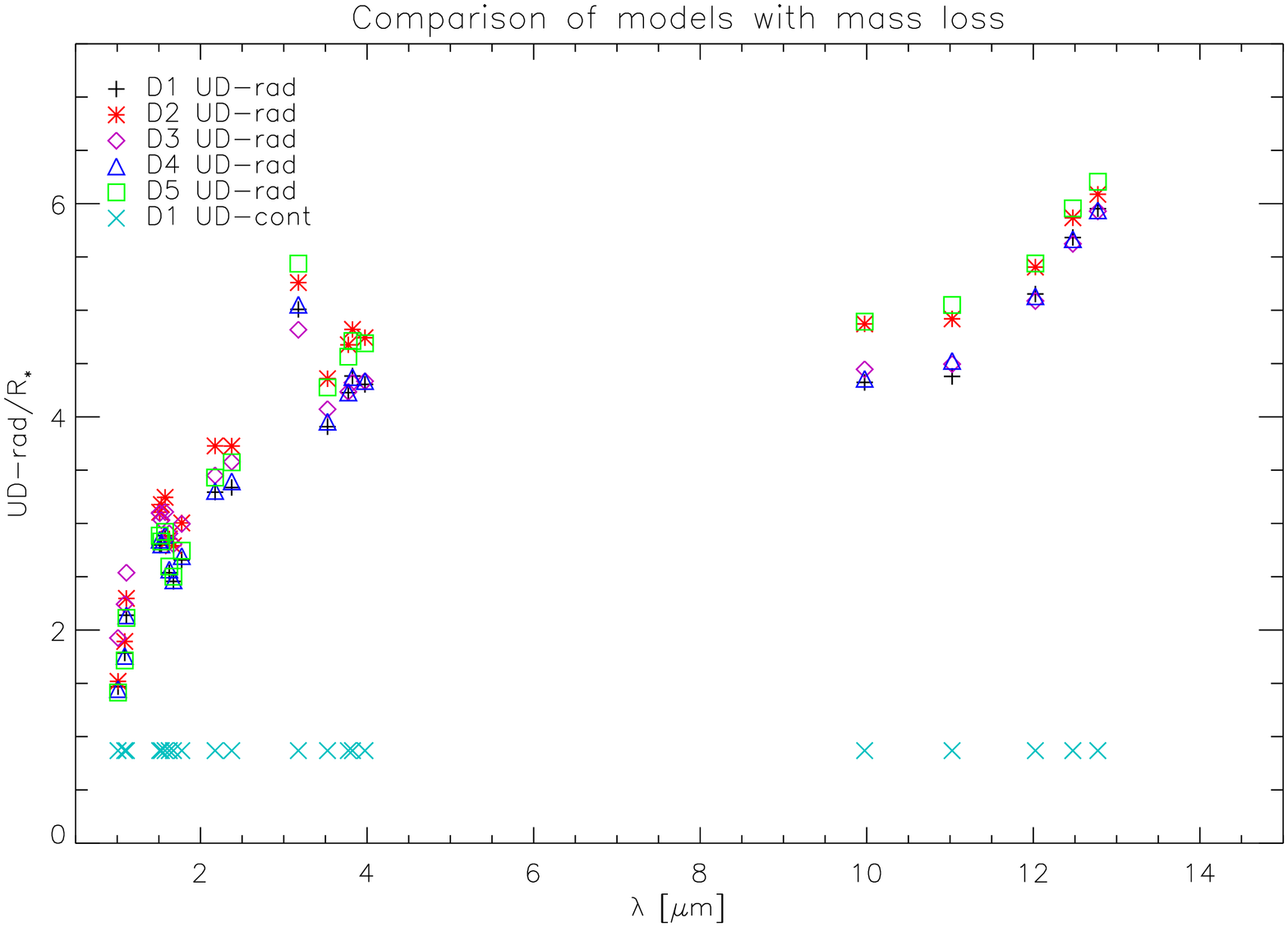}\\
	   \includegraphics[angle = 90, width = \hsize]{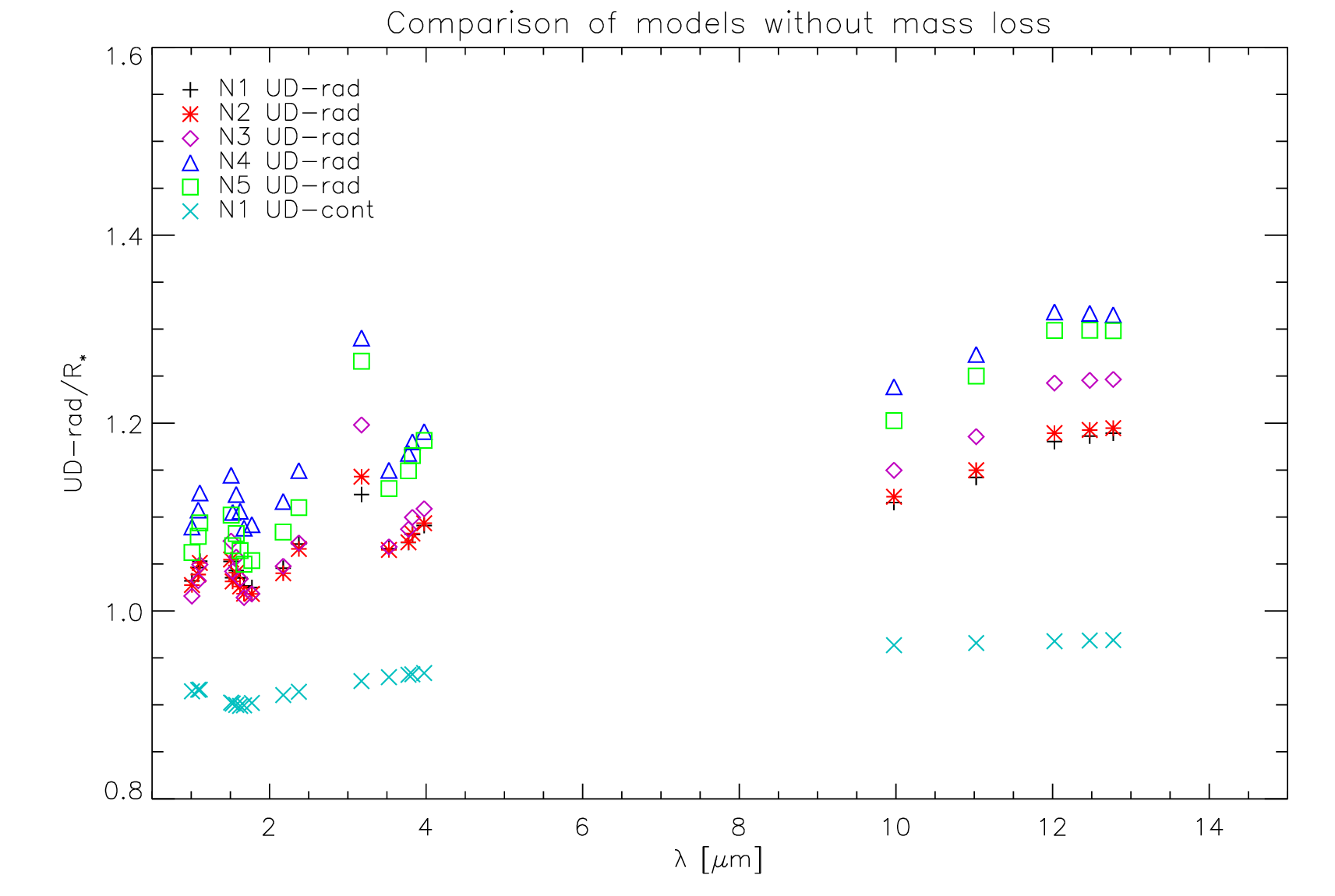}
	   \caption{UD-radii versus wavelength for the maximum phase of different models 
	 with (upper panel) and without (lower panel) mass loss. The cross symbol represents the
         UD-continuum radius computed for the two reference models (D1, N1).
	 The method used for the UD determination is a least-square fit using all 
	 the points of the synthetic profile with visibility $> 0.1$.}
	              \label{rlambda.fig}
	    \end{figure}
        Comparing the effect of the different model parameters we can say that model extensions increase with
        decreasing effective temperature and higher \mbox{mass loss} values.
        The models D1 and D2 for example are distinguished by a higher \mbox{mass loss} rate for D2 because of the different period
        and $f_{\rm{L}}$ value.
        The resulting UD-radii for D2 are systematically higher than the one for D1.
        D5 has a higher piston velocity, and consequently a higher \mbox{mass loss}, compared to D4.
        The UD-radii obtained are larger.
        A comparison of D1-D2 or D4-D5 in Fig.~\ref{rlambda.fig} 
        shows that the effect of the \mbox{mass loss} rate on the UD-radius is larger
        beyond $3\,\mu$m.
        The model D3 has UD-radii systematically larger than the one of the model D4 in the region between 1 and 2 $\mu$m and at 9 $\mu$m.
        At 3.175 $\mu$m the trend is inverted while for the other filters at 3 $\mu$m and for the longer wavelength the resulting UD-radii 
        is the same for the two models.
        D3 and D4 differ in the temperature (2\,600 K for D3 and 2\,800 K for D4), piston velocity, period and
        $f_{\rm{L}}$ while the resulting \mbox{mass loss} rate is quite similar.
        
        On the other hand, in absence of dusty winds the models with the same parameters except for C/O 
        (N2 versus N3) have very similar radii.
        The difference is barely detectable at 3 $\mu$m
        and at longer wavelengths. 
        If, besides different C/O ratios, also the temperature of the model changes (e.g. as between N4 and N5)
        the resulting UD-radii show more pronounced differences. 
        The UD-radii of N1 overlap always with the one of N2, even if the models have different temperature (higher for N1), luminosity
        (higher for N2) and period (longer for N2).
        N3 and N4 have all the same parameters except the piston velocity and the resulting UD-radii is
        larger for the second model, which has higher value of $\Delta u_{\rm{p}}$.
        Model N4 is also the most extended among its subgroup.

        \subsection{Separating models observationally
        }

        The dynamic model atmospheres provide, compared with classical hydrostatic models, 
        a self-consistent treatment of time-dependent phenomena (shock propagation, 
        stellar wind, dust formation). It is thus of course highly interesting if observed UD-radii at different wavelengths
        can be used to distinguish between the different models and then to assign a given model to a given star.

        If the model UD-radii are {\em not} scaled to their respective $R_\star$, the D models differ by $30\%$ to $50\%$. These 
        differences should be detectable given typical distance uncertainties of $20\%$. However, as dicussed above, this would mostly 
        separate effects due to the fundamental stellar parameters (initial radius, luminosity) and allow to discriminate between N and D models
        but dynamic effects within each series are only of the order of $10\%$ to $25\%$. This can be seen from Fig.~\ref{rlambda.fig} or
        if one scales each model to e.g.\ its UD-radius in the K-band. The latter approach would be the most sensitive one as it eliminates
        errors in the distance (provided the measurements were done at the same pulsational phase, better at the same time!). 
        Assuming that individual UD-radii can be determined with an accuracy of $5\%$ to $10\%$, the dynamic effects 
        could be (barely) distinguishable for the D models. The most sensitive wavelengths in this respect would be
        filters in the $K$- and $L$-band. The $N$-band shows a similar sensitivity to dynamic effects though at a reduced observational accuracy.
        For the N models these effects are less then 10\%, with the largest difference again beyond the $K$-band.

        Nevertheless, the best strategy for choosing the right dynamic model is to combine spectroscopic
        and interferometric observations. 
        
        Finally it should be noted, that for a larger grid of models \citep[e.g.][]{mat08} a wider separation of models might show up.
        This will however be the subject of a future paper.

	\begin{figure*}
	\centering
	 \includegraphics[angle = 90,width = 17cm]{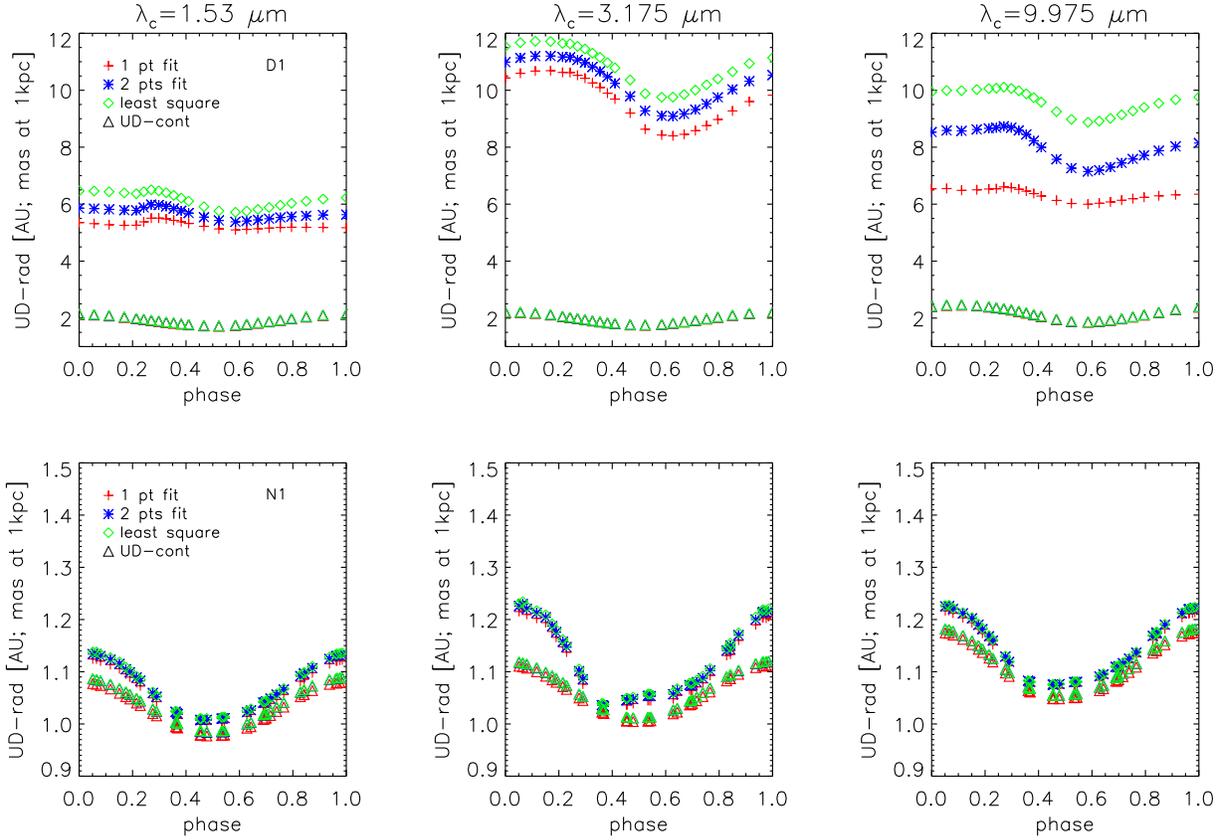}
	 \caption{ UD-radii versus phase are shown for three different narrow-band filters. 
           The three upper panels represent 
	results for the dynamic model D1 with mass loss, the three panels at the bottom are 
	for the model N1 without mass loss. 
        Triangles are for UD-radius computed taking into account only continuum opacity. 
        For the continuum the same symbol is used for the different methods because there is no appreciable difference between the results.
        Crosses, stars and diamonds represent the different methods (one point, two points, least-square, respectively) 
        used for fitting the visibility profiles of models with UDs.} 
        \label{rphase.fig}
	\end{figure*}

        \subsection{UD-radii as a function of time}
        The behaviour of the UD-radius versus phase predicted by the models is rather complex, 
        especially in the case of mass-losing models.
        In Fig.~\ref{rphase.fig} the UD-radius at a given wavelength/filter versus phase is plotted for the two reference models.
        The presence of molecular and dust opacities increases the UD-radius strongly compared to the UD-continuum radius. 
	In both types of models we can observe the periodic movement of the stellar interior, which is pronounced to
 	a different degree for different models of Table~\ref{mod.tbl}.
        The periodic variation of the UD-radius in all the models without \mbox{mass loss} reflects 
        the fact that the movement of the complete atmosphere follow mostly the pulsation.
        The time scales introduced by dust-formation in the atmosphere of the models with \mbox{mass loss} may deform the sinusoidal 
        pattern, a behaviour that can be explained with outer layers which are not
        moving parallel to the inner ones.
        This is well represented in Fig.~2 of \citet{hoe03}, but one should keep in mind that the movement of theoretical mass shells 
        should not be directly compared to the changes in the atmospheric radius-density structure
        that we observe by interferometric measurements.
        
        From the lower three panels of Fig.~\ref{rphase.fig} 
        (no \mbox{mass loss}) it is obvious that there is not much difference between the UD-radius 
        and UD-continuum radius for the model N1, especially for the 1.53 $\mu$m and 9.975 $\mu$m filter 
        close to minimum luminosity. We find the largest difference for the maximun luminosity ($\phi = 0$)
        where the ratio between UD-radius and UD-continuum radius is 1.09 for the filter 3.175 $\mu$m 
        and 1.04 for the other two filters.
        For the model D1 (upper panels) the value of the same ratio is 
        3.03 for the filter 1.53 $\mu$m, 4.29 for 3.175 $\mu$m and 4.12
        for 9.975 $\mu$m, much larger than the values found for the models without mass-loss.

	The three different methods of fitting 
        give mostly the same result in the case of the model without mass loss, 
	while in the case of models with mass loss, the UD-radius is smaller for the 
        first 2 methods (one- and two-points fit) than for the 
        least square result.
	This last result can be explained going back to our initial classification of profiles (Sect.~\ref{synth.sect}). 
        The first and the third class of profiles belong to the ``uncommon'' profiles of S03, not approximable with UD functions.
        In particular the third class of profiles (9.975 $\mu$m)
        is the one that mostly deviates from a UD (see Fig.~\ref{prof.fig}).

        Even increasing the number of fitting points, 
        the resulting UD-radius cannot be considered reliable due to the particular shape of the profiles. 
        From Fig.~\ref{rphase.fig} it is also clear that for the narrow-band filters it is not possible to define a 
        single, phase independent scaling factor that gives the possibility to convert from UD-radii to UD-continuum radii.

        \begin{figure*}
	\centering
	 \includegraphics[angle = 90,width = 17cm]{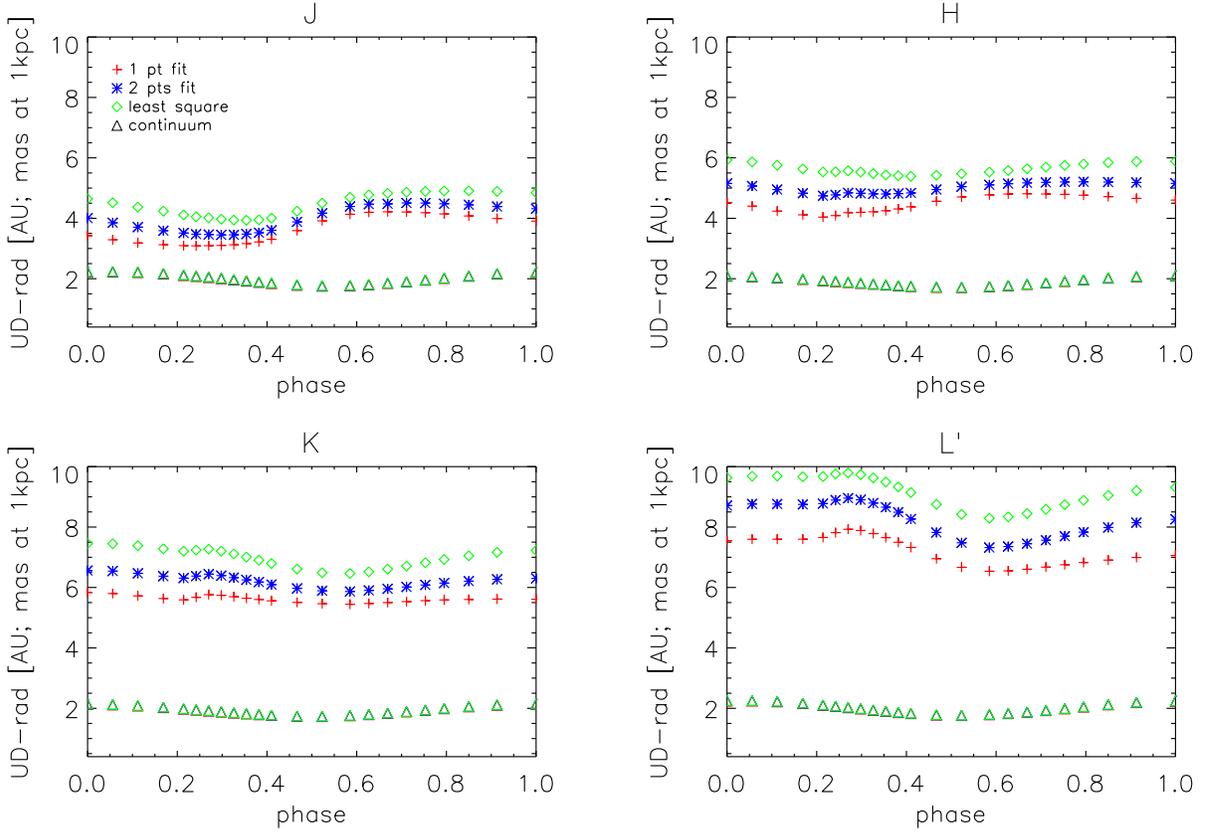}
         \caption{\small UD-Radius versus phase for $JHKL^\prime$ broad-band filters. 
           The four panels show the results for the model D1 with \mbox{mass loss}.Plot symbols are the same as in Fig.~\ref{rphase.fig}.}
        \label{rphasebroad.fig}
	\end{figure*}
        
        \begin{figure*}
	\centering
        \includegraphics[angle = 90,width = 17cm]{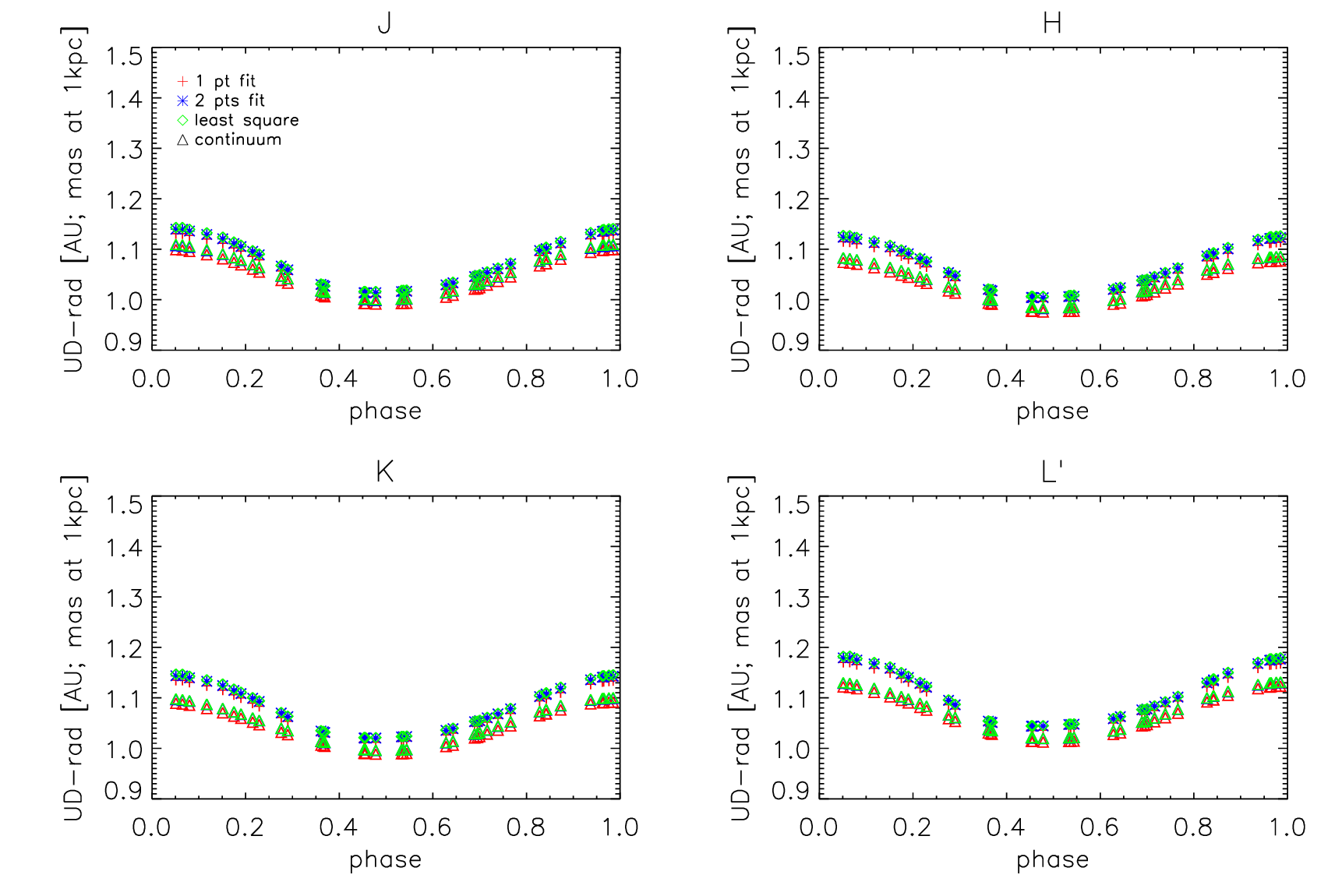}
        \caption{\small UD-Radius versus phase for $JHKL^\prime$ broad-band filters. 
           The four panels show the results for the model N1 without \mbox{mass loss}.  Plot symbols the same as in Fig.~\ref{rphase.fig}.}
       \label{rphasebroad2.fig}
	\end{figure*}

        We also computed the equivalent UD-radii and UD-continuum radii for the broad-band filters. 
        The results are shown in  Fig.~\ref{rphasebroad.fig}.  
        The broad-band has the effect  of smoothing the visibility profile and
        this can be noted in the UD-radii behaviour versus phase by the reduced amplitude of the radius-variation.
        As for the narrow-band also in the case of broad-band filters 
        the UD-radius versus phase for the set of models D shows a behaviour
        not synchronized with the pulsation.
        The ratio between UD-radius and UD-continuum radius is not constant with phase and/or with wavelength.
        For the phase zero it corresponds to 2.08 in $J$, 
        2.83 in $H$, 2.14 in $K$ and 2.54
        in $L^\prime$ band.
        In the lower panels of Fig.~\ref{rphasebroad.fig}, where the model N1 is represented, the resulting UD-radius 
        is always closer to the UD-continuum radius. The ratio between UD-radius and UD-continuum radius for the phase zero is 
        almost constant: 1.03 for $J$ and $H$,1.04 for $K$ and $L^\prime$.
        Again, the amplitudes of variation of the radius for the N models are smaller than for the narrow-band filters, and 
        the variation follows the near-sinusoidal behaviour of the inner layers.
        
        For the broad-band filters the difference in the fitting methods 
        confirms the trend found for the narrow-band case.
        For the models with \mbox{mass loss}
        the one-point and two-points fit methods give smaller UD-radii compared with the one resulting 
        from the least square method.
        The UD-radii, resulting from different methods, agree closely in the case of models without \mbox{mass loss}
        and the same result is obtained also for all the UD-continuum measurements.
        
        The dependence of UD-radius on the pulsation cycle is not covered by this contribution, 
        it will be topic of a future contribution.

        \subsection{Comparison with \mbox{M-type} stars}
        \label{mstars.sect}

        Compared to the \mbox{M-type} stars
        we emphasise the lack of a spectral window for measuring a pure continuum radius. 
        A model-dependent definition of the continuum radius is thus needed for the cool Carbon stars.
        In \citet{ire04a} the phase variation of filter radii in units of the parent star radius $R_{{p}}$\footnote{$R_{{p}}$ is defined as
        the Rosseland radius that ``the Mira variable would assume if the pulsation stopped and 
        the stratification of the star becomes static'' \citep{jac02}.}
        is represented for different near-continuum narrow and broad-band filters.
        They demonstrate how measurements of radii in near-continuum band-passes are affected by 
        molecular contaminations that mask the geometrical pulsation. This contamination is
        negligible only close
        to the visual maximum. Other suitable band-passes for measuring the pulsation of the continuum layers
        for \mbox{M-type} stars are the broad-band filters $J$ and $H$, or sub-bandpasses of $H$ and $K$.
        
        To carry out a consistent comparison between the UD-radii computed in this work for C-stars, and the resulting UD-radii of \citet{jac02}
        for \mbox{M-type} stars is difficult because of the intrinsc difference in the modelling approach, and because of the different stellar
        parameters. But we can say as general statement that the ratios of UD-radii/UD-continuum that we obtain
        are systematically larger for our models with \mbox{mass loss} than the one presented for \mbox{M-type} stars in Figs.~4 and ~5 of \citet{ire04a}.
        More comparable are the resulting ratios for the models without \mbox{mass loss}.        
        The largest difference is due to the strong molecular (first of all C$_2$H$_2$) and dust opacities
        that characterize the C-star spectra. 
        The UD-radius increases with wavelength and in the case of Carbon-rich model atmospheres with \mbox{mass loss}
        this increase is larger than the one found by \citet{jac02} for \mbox{M-type} stars. 
        In particular, the ratio between UD-radius in $J$ and UD-radius in $L^\prime$ for our model D1 at phase zero is
        2.07, while for the \mbox{M-type} models presented in Fig.~4 and Fig.~5 of \citet{jac02} is 1.4 in the case of largest discrepancy (phase 1.5).
        The same ratio for our N1 model is 1.03, meaning that this model without \mbox{mass loss} shows an increase of the UD-radius with wavelength
        lower than the one of the \mbox{M-type} stars.
                
	\section{Conclusions}
        In this paper we present, for the first time, a theoretical study on the intensity and visibility profiles computed for a set
        of dynamic model atmospheres of \mbox{C-rich} AGB stars.
        The main results of our investigation can be summarised as follows:
	\begin{itemize}
        \item The profiles computed in the narrow-band filters for our set of models can be divided in 3 morphological classes 
          mainly according to wavelength region, with the exception of the filter located at 3.175 $\mu$m 
          in the middle of the strong absorption feature characteristic for C-star spectra.
          This filter shows a behaviour
          more closely related to the one of the narrow-band filters in the region around 12 $\mu$m 
          than to the surrounding wavelength range.
          We have chosen three filters to represent the morphology of intensity and visibility profiles: 1.53, 9.975 and $3.175\,\mu$m 
          respectively for the first, second and third class.
          The intensity profile of the first class is characterized by a limb darkened disc followed by a tail-shape with a bump and a peak. 
          The $3.175$-like profiles show an extended plateau due to strong molecular opacity. The second class is intermediate between the other two.   
        \item The morphology of the profiles (narrow and broad-band) of models
          with \mbox{mass loss} is very far from the UD shape. Models without \mbox{mass loss}, and all the
          theoretical continuum profiles can be quite well approximated by a UD.
        \item Bandwidth smearing affects the resulting broad-band profiles with an error of about 70\% for certain baselines.
          For models with \mbox{mass loss} the difference is notable not only close to the first zero but also on the flank of the
          first lobe of the visibility profile. For a comparison of models with observations in broad-band it is important to 
          take this effect properly into account.           
	\item C-star UD-radii are wavelength-dependent. The dependency is stronger in the case of models with \mbox{mass loss}, in this case the UD-radii
          show also large phase dependence.
	Around $3\,\mu$m and in the N-band the star appears more extended due to C$_2$H$_2$ opacity. The UD-continuum radius is mostly independent
        of the wavelength and phase.
        \item The extensions of the models increase for higher \mbox{mass loss} rate and lower temperature. 
        \item Models with \mbox{mass loss} show a complicated behaviour of the UD-radius versus phase compared to models without \mbox{mass loss}. 
          This behaviour does not follow the pulsation of the interior. 
	\item It is shown that using only one or two points of visibility to determine the UD-radii of objects characterised by
          strong pulsation and \mbox{mass loss}, the radius of the star is smaller than the one measured using more points (least square method). 
          In the case of models without \mbox{mass loss} 
          the different methods of fitting give the same UD-radii.
	\item The UD-radius is very close to the UD-continuum radius in the case of models without \mbox{mass loss}.
        \item The trend of the UD-radii versus phase in broad-band is smoother than in the case of the narrow-band filters.
        \item Contrary to O-rich stars, for \mbox{C-rich} stars
          no spectral windows for observing the continuum radius are available in the 
          case of C-stars and some
          assumptions based on models are necessary. 
          A useful replacement for a measured continuum radius is the UD-continuum radius defined in Sect.~\ref{ud.sect}
          which is basically independent of wavelength and pulsation phase.
        \item The difference in the UD-radii between the models should be detectable observationally, in particular in the $L$-band
        where MATISSE \citep{lop06} will provide new opportunity.
	\item The radius computed with the UD function has to be condsidered only as a first guess for the real size of the star.
	The intensity and the visibility profiles of a C-star, especially in the case of models producing \mbox{mass loss}, 
        are very far from being UD-like.  

	\end{itemize}
        A comparison of the dynamic models with available sets of observations is under way, intensity and visibility profiles 
        for specific models can be provided upon request.         
	
	\begin{acknowledgements}
	 This work has been supported by Projects P19503-N16 and P18939-N16 of the Austrian Science Fund (FWF). It is a pleasure to thank T.
         Verhoelst and T. Driebe for many helpful comments and suggestions. BA acknowledges funding by the contract ASI-INAF I/016/07/0.
	\end{acknowledgements}

	\end{document}